\documentclass[10pt,preprint]{aastex}
\usepackage{psboxit}
\usepackage{lscape}

\PScommands
\newcommand{\graybox}[1]{\psboxit{box 0.7 setgray fill}{\spbox{#1}}}

\newcommand{\Sersic}{S\'ersic}

\newcommand{\avg}[1]{{\langle{#1}\rangle}}
\newcommand{\Avg}[1]{{\left\langle{#1}\right\rangle}}
\def\simless{\mathbin{\lower 3pt\hbox
	{$\,\rlap{\raise 5pt\hbox{$\char'074$}}\mathchar"7218\,$}}} % < or of order
\def\simgreat{\mathbin{\lower 3pt\hbox
	{$\,\rlap{\raise 5pt\hbox{$\char'076$}}\mathchar"7218\,$}}} % > or of order

\newcommand{\petroratio}{{{\mathcal{R}}_P}}
\newcommand{\petroradius}{{{\theta}_P}}

\newcommand{\petroratiolim}{{{\mathcal{R}}_{P,\mathrm{lim}}}}
\newcommand{\band}[2]{\ensuremath{^{#1}\!{#2}}}

\newcounter{thefigs}
\newcommand{\fignum}{\arabic{thefigs}}

\newcounter{thetabs}

\newcounter{address}

\slugcomment{Submitted to \apj}

\shortauthors{Blanton {\it et al.} (2002)}
\shorttitle{Galaxy Luminosity Density at $z=0.1$}

\begin{document}
 

\title{The Galaxy Luminosity Function and Luminosity Density \\
at Redshift $z=0.1$\altaffilmark{\ref{SDSS}}}


%\author{Michael Blanton}
%\affil{NASA/Fermilab Astrophysics Center\\
%Fermi National Accelerator Laboratory, Batavia, IL 60510-0500}
%\author{\and lots and lots of other, probably more important people}
%\affil{Some Institution\\
%Somewhere, Some City, Some State or Province}
%\email{blanton@fnal.gov}

% Authorship determined by those I was directly involved with
% in performing this work as well as those responsible for the photo-z
% plates and those who have characterized the filter curves (since
% these might be publicly distributed). 
\author{
Michael R. Blanton\altaffilmark{\ref{NYU}},
David W. Hogg\altaffilmark{\ref{NYU}},
%Tamas Budavari\altaffilmark{\ref{JHU}},
J.~Brinkmann\altaffilmark{\ref{APO}},
Andrew J. Connolly\altaffilmark{\ref{Pitt}},
Istv\'an Csabai\altaffilmark{\ref{JHU}},
%Mamoru Doi\altaffilmark{\ref{Tokyo}},
%Daniel Eisenstein\altaffilmark{\ref{Arizona}},
%James E. Gunn\altaffilmark{\ref{Princeton}},
%David W. Hogg\altaffilmark{\ref{NYU}}, and
%David J. Schlegel\altaffilmark{\ref{Princeton}}
%Julianne Dalcanton\altaffilmark{\ref{UW}},
%Mark SubbaRao\altaffilmark{\ref{Chicago}},
%John E. Anderson, Jr.\altaffilmark{\ref{Fermilab}},
%James Annis\altaffilmark{\ref{Fermilab}},
Neta A. Bahcall\altaffilmark{\ref{Princeton}},
%Mariangela Bernardi\altaffilmark{\ref{Chicago}},
%Robert J. Brunner\altaffilmark{\ref{Caltech}},
%Scott Burles\altaffilmark{\ref{Fermilab}},
%Larry Carey\altaffilmark{\ref{UW}},
%Francisco J. Castander\altaffilmark{\ref{Chicago}, \ref{Pyrenees}},
%Andrew J. Connolly\altaffilmark{\ref{Pitt}},
%Istv\'an Csabai\altaffilmark{\ref{JHU}},
%Douglas Finkbeiner\altaffilmark{\ref{Berkeley}},
%Scott Friedman\altaffilmark{\ref{JHU}},
%Joshua A. Frieman\altaffilmark{\ref{Fermilab}},
%G. S. Hennessy\altaffilmark{\ref{USNO}},
%Robert B. Hindsley\altaffilmark{\ref{USNO}},
%Takashi Ichikawa\altaffilmark{\ref{Tokyo}},
%\v{Z}eljko Ivezi\'{c}\altaffilmark{\ref{Princeton}},
%Stephen Kent\altaffilmark{\ref{Fermilab}},
%G. R.~Knapp\altaffilmark{\ref{Princeton}},
%D. Q.~Lamb\altaffilmark{\ref{Chicago}},
%R. French Leger\altaffilmark{\ref{UW}},
%Daniel C. Long\altaffilmark{\ref{APO}},
%Robert H. Lupton\altaffilmark{\ref{Princeton}},
%Timothy A.~McKay\altaffilmark{\ref{Michigan}},
Masataka Fukugita\altaffilmark{\ref{CosmicRay}},
Jon Loveday\altaffilmark{\ref{Sussex}},
Avery Meiksin\altaffilmark{\ref{Edinburgh}},
%Aronne Merelli\altaffilmark{\ref{Caltech}},
Jeffrey A. Munn\altaffilmark{\ref{USNO}},
%Vijay Narayanan\altaffilmark{\ref{Princeton}},
%Matt Newcomb\altaffilmark{\ref{CarnegieMellon}},
R. C. Nichol\altaffilmark{\ref{CarnegieMellon}},
Sadanori Okamura\altaffilmark{\ref{Tokyo}},
%Russell Owen\altaffilmark{\ref{UW}},
%Jeffrey R.~Pier\altaffilmark{\ref{USNO}},
%Adrian Pope\altaffilmark{\ref{JHU}},
%Marc Postman\altaffilmark{\ref{STScI}},
Thomas Quinn\altaffilmark{\ref{UW}},
%Constance M. Rockosi\altaffilmark{\ref{Chicago}},
Donald P. Schneider\altaffilmark{\ref{PennState}}, 
Kazuhiro Shimasaku\altaffilmark{\ref{Tokyo}},
Michael A. Strauss\altaffilmark{\ref{Princeton}},
Max Tegmark\altaffilmark{\ref{UPenn}},
%Walter A. Siegmund\altaffilmark{\ref{UW}},
%Stephen Smee\altaffilmark{\ref{Maryland}},
%Yehuda Snir\altaffilmark{\ref{CarnegieMellon}},
%Chris Stoughton\altaffilmark{\ref{Fermilab}},
%Christopher Stubbs\altaffilmark{\ref{UW}},
%Alexander S.~Szalay\altaffilmark{\ref{JHU}},
%Gyula P.~Szokoly\altaffilmark{\ref{Potsdam}},
%Aniruddha R.~Thakar\altaffilmark{\ref{JHU}},
%Christy Tremonti\altaffilmark{\ref{JHU}},
%Douglas L. Tucker\altaffilmark{\ref{Fermilab}},
%Alan Uomoto\altaffilmark{\ref{JHU}},
%Dan vanden Berk\altaffilmark{\ref{Fermilab}},
Michael S. Vogeley\altaffilmark{\ref{Drexel}}, and
David H. Weinberg\altaffilmark{\ref{Ohio}}
%Patrick Waddell\altaffilmark{\ref{UW}},
%Brian Yanny\altaffilmark{\ref{Fermilab}},
%Naoki Yasuda\altaffilmark{\ref{NAOJ}},
%and Donald G.~York\altaffilmark{\ref{Chicago}}
\vspace{3.0in}
}

\altaffiltext{1}{Based on observations obtained with the
Sloan Digital Sky Survey\label{SDSS}} 
\setcounter{address}{2}
\altaffiltext{\theaddress}{
\stepcounter{address}
Center for Cosmology and Particle Physics, Department of Physics, New
York University,  4 Washington Place, New York, NY 10003
\label{NYU}}
\altaffiltext{\theaddress}{
\stepcounter{address}
Apache Point Observatory,
2001 Apache Point Road,
P.O. Box 59, Sunspot, NM 88349-0059
\label{APO}}
\altaffiltext{\theaddress}{
\stepcounter{address}
University of Pittsburgh,
Department of Physics and Astronomy,
3941 O'Hara Street,
Pittsburgh, PA 15260
\label{Pitt}}
\altaffiltext{\theaddress}{
\stepcounter{address}
Department of Physics and Astronomy, The Johns Hopkins University,
Baltimore, MD 21218
\label{JHU}}
\altaffiltext{\theaddress}{
\stepcounter{address}
Princeton University Observatory, Princeton,
NJ 08544
\label{Princeton}}
\altaffiltext{\theaddress}{
\stepcounter{address}
Institute for Cosmic Ray Research, University of
Tokyo, Midori, Tanashi, Tokyo 188-8502, Japan
\label{CosmicRay}}
\altaffiltext{\theaddress}{
\stepcounter{address}
Sussex Astronomy Centre,
University of Sussex,
Falmer, Brighton BN1 9QJ, UK
\label{Sussex}}
\altaffiltext{\theaddress}{
\stepcounter{address}
Department of Physics \& Astronomy,
The University of Edinburgh,
James Clerk Maxwell Building,
The King's Buildings,
Mayfield Road,
Edinburgh EH9 3JZ, UK
\label{Edinburgh}}
\altaffiltext{\theaddress}{
\stepcounter{address}
U.S. Naval Observatory,
3450 Massachusetts Ave., NW,
Washington, DC  20392-5420
\label{USNO}}
\altaffiltext{\theaddress}{
\stepcounter{address}
Department of Physics, Carnegie Mellon University, 
5000 Forbes Avenue, Pittsburgh, PA 15213-3890 
\label{CarnegieMellon}}
\altaffiltext{\theaddress}{
\stepcounter{address}
Department of Astronomy and Research Center for 
the Early Universe,
School of Science, University of Tokyo,
Tokyo 113-0033, Japan
\label{Tokyo}}
\altaffiltext{\theaddress}{
\stepcounter{address}
Department of Astronomy, University of Washington,
Box 351580,
Seattle, WA 98195 
\label{UW}}
\altaffiltext{\theaddress}{
\stepcounter{address}
Department of Astronomy and Astrophysics,
The Pennsylvania State University,
University Park, PA 16802
\label{PennState}}
\altaffiltext{\theaddress}{
\stepcounter{address}
Department of
Physics and Astronomy,
209 South 33rd Street,
Philadelphia, PA 19104-6396
\label{UPenn}}
\altaffiltext{\theaddress}{
\stepcounter{address}
Department of Physics, Drexel University, Philadelphia, PA 19104
\label{Drexel}}
\altaffiltext{\theaddress}{
\stepcounter{address}
Ohio State University,
Department of Astronomy,
Columbus, OH 43210
\label{Ohio}}

\clearpage

\begin{abstract}
Using a catalog of 147,986 galaxy redshifts and fluxes from the Sloan
Digital Sky Survey (SDSS) we measure the galaxy luminosity density at
$z=0.1$ in five optical bandpasses corresponding to the SDSS
bandpasses shifted to match their restframe shape at $z=0.1$. We
denote the bands $\band{0.1}{u}$, $\band{0.1}{g}$, $\band{0.1}{r}$,
$\band{0.1}{i}$, $\band{0.1}{z}$, with $\lambda_{\mathrm{eff}} =
[3216$, 4240, 5595, 6792, 8111 \AA$]$ respectively. To estimate the
luminosity function, we use a maximum likelihood method which allows
for a general form for the shape of the luminosity function, simple
luminosity and number evolution, incorporates the flux uncertainties,
and accounts for the flux limits of the survey. We find luminosity
densities at $z=0.1$ expressed in absolute AB magnitudes in a Mpc$^3$
to be $[-14.10 \pm 0.15, -15.18 \pm 0.03, -15.90 \pm 0.03, -16.24 \pm
0.03, -16.56 \pm 0.02]$ in $[$\band{0.1}{u}, \band{0.1}{g},
\band{0.1}{r}, \band{0.1}{i}, \band{0.1}{z}$]$, respectively, for a
cosmological model with $\Omega_0 =0.3$, $\Omega_\Lambda=0.7$, and
$h=1$, and using SDSS Petrosian magnitudes. Similar results are
obtained using \Sersic\ model magnitudes, suggesting that flux from
outside the Petrosian apertures is not a major correction. In the
$\band{0.1}{r}$ band, the best fit Schechter function to our results
has $\phi_\ast = (1.49 \pm 0.04) \times 10^{-2}$ $h^3$ Mpc$^{-3}$,
$M_\ast - 5\log_{10} h = -20.44 \pm 0.01$, and $\alpha = -1.05\pm
0.01$. In solar luminosities, the luminosity density in \band{0.1}{r}
is $(1.84 \pm 0.04)$ $h$ $10^8$ $L_{\band{0.1}{r},\odot}$
Mpc$^{-3}$. Our results are consistent with other estimates of the
luminosity density, from the Two-degree Field Galaxy Redshift Survey
and the Millenium Galaxy Catalog. They represent a substantial change
($\sim 0.5$ mag) from earlier SDSS luminosity density results based on
commissioning data, almost entirely because of the inclusion of
evolution in the luminosity function model.
\end{abstract}


% TODO:  
%
% Later:
% QA plot on log scale?
% mention deviations at bright end
% more explicit rejection of fitting Schechter models

%
% Introduction and motivation
%

\section{Motivation}
\label{motivation}

Establishing the low-redshift galaxy luminosity density of the
universe is a fundamental measurement of the contents of the local
universe. The last two decades, beginning with the Center for
Astrophysics redshift survey (\citealt{huchra83a}), have seen steady
progress in understanding the total galaxy luminosity density emitted
in the universe. Part of the progress has been due to measuring larger
and larger numbers of galaxy fluxes and redshifts, from only a few
hundred redshifts before 1980 to a few hundred thousand at
present. However, equally importantly, the determination of galaxy
fluxes has been steadily improving. The luminosity functions
calculated from the CfA survey were based on ``Zwicky'' magnitudes:
essentially, they were determined by visual inspection of photographic
material.  Automated processing of photographic plates provided an
improved way of measuring flux; the Automatic Plate Measurement survey
(\citealt{maddox90a}), from which targets were selected for the
Two-degree Field Galaxy Redshift Survey (2dFGRS;
\citealt{colless01a}), is the latest example of this method. However,
deep CCD imaging, such as that performed by the Sloan Digital Sky
Survey (SDSS; \citealt{york00a}), which provides a higher dynamic
range and more linear response than do photographic plates, is
yielding the highest signal-to-noise ratio measures of flux and the
greatest surface brightness sensitivity in a large survey to date.

A preliminary estimate of the SDSS galaxy luminosity function was
performed by \citet{blanton01a}, using a small sample of commissioning
data.  At the time of writing, the SDSS has obtained photometry and
redshifts for more than ten times the number of galaxies in the
commissioning data used by \citet{blanton01a}; furthermore, the
photometric calibration procedures have improved since that time.  In
addition, we have developed a new and more self-consistent method for
$K$-correcting galaxies to a fixed frame bandpass
(\citealt{blanton02b}). Finally, we have developed a maximum
likelihood technique for fitting the galaxy luminosity function which
allows for generic luminosity function shapes as well as luminosity
and number density evolution (\citealt{blanton02e}). Thus, it is now
time to reevaluate the galaxy luminosity function in the SDSS.

Throughout this paper we assume a Friedmann-Robertson-Walker
cosmological world model with matter density $\Omega_0 = 0.3$, vacuum
pressure $\Omega_\Lambda = 0.7$, and Hubble constant $H_0 = 100$
$h$ km s$^{-1}$ Mpc$^{-1}$ with $h=1$, unless otherwise specified.

In Section \ref{data}, we describe the SDSS galaxy catalog data. In
Section \ref{method}, we briefly describe our method of fitting the
luminosity function (a fuller description of the method can be found
in \citealt{blanton02e}). In Section \ref{results}, we present our
results for the galaxy luminosity density at $z=0.1$. In Section
\ref{systematics}, we test whether our results are sensitive to how we
determine the galaxy magnitudes. In Section \ref{comparison}, we
compare our results to previous work. We conclude and discuss plans
for future work in Section \ref{conclusions}.

\section{SDSS Spectroscopic Galaxy Catalog}
\label{data}

The SDSS (\citealt{york00a}) is producing imaging and spectroscopic
surveys over $\pi$ steradians in the Northern Galactic Cap. A
dedicated 2.5m telescope (Siegmund et al., in preparation) at Apache
Point Observatory, Sunspot, New Mexico, images the sky in five bands
between 3000 and 10000 \AA\ ($u$, $g$, $r$, $i$, $z$;
\citealt{fukugita96a}) using a drift-scanning, mosaic CCD camera
(\citealt{gunn98a}), detecting objects to a flux limit of $r\sim 22.5$
mags. A major goal of the survey is to spectroscopically observe
900,000 galaxies (down to $r_{\mathrm{lim}}\approx 17.77$ mags),
100,000 Luminous Red Galaxies (\citealt{eisenstein01a}), and 100,000
QSOs (\citealt{fan99a}, \citealt{richards02a}) selected from the
imaging data. This spectroscopic follow up uses two digital
spectrographs on the same telescope as the imaging camera. Many of the
details of the galaxy survey are described in the galaxy target
selection paper (\citealt{strauss02a}). Other aspects of the survey
are described in the Early Data Release paper (EDR;
\citealt{stoughton02a}). The survey has begun in earnest, and has so
far obtained about 30\% of its intended data.

\subsection{SDSS Observational Analysis}

The SDSS images are reduced and catalogs are produced by the SDSS
pipeline {\tt photo}, which measures the sky background and the local
seeing conditions, and detects and measures objects.  The astrometric
calibration is performed by an automatic pipeline which obtains
absolute positions to better than 0.1 arcsec (\citealt{pier02a}).  The
magnitudes are calibrated to a standard star network
(\citealt{smith02a}) approximately in the AB system. There are small
differences between the system output by the SDSS pipelines and a true
AB system, amounting to $\Delta m = -0.042, 0.036, 0.015, 0.013,
-0.002$ in the $u$, $g$, $r$, $i$, and $z$ bands. Because these
effects were discovered at a relatively late date in the preparation
of this manuscript, we have not self-consistently included these
shifts in our results. Instead we have applied them {\it a posteriori}
to our results in the $\band{0.1}{u}$, $\band{0.1}{g}$,
$\band{0.1}{r}$, $\band{0.1}{i}$, and $\band{0.1}{z}$ bands.

Object fluxes are determined several different ways by {\tt photo}, as
described in \citet{stoughton02a}. The primary measure of flux used
for galaxies is the SDSS Petrosian magnitude, a modified version of
the quantity proposed by \citet{petrosian76a}. The essential feature
of Petrosian magnitudes is that in the absence of seeing they measure
a constant fraction of a given galaxy's light regardless of distance
(or size).  More specifically, we define the ``Petrosian ratio''
$\petroratio$ at an angular radius $\theta$ from the center of an
object to be the ratio of the local surface brightness averaged over
an annulus at $r$ to the mean surface brightness within $\theta$:
\begin{equation}
\label{petroratio}
\petroratio (\theta)\equiv \frac{\left.
\int_{\alpha_{\mathrm{lo}} \theta}^{\alpha_{\mathrm{hi}} \theta} d\theta' 2\pi \theta'
I(\theta') \right/ \left[\pi(\alpha_{\mathrm{hi}}^2 -
\alpha_{\mathrm{lo}}^2) \theta^2\right]}{\left.
\int_0^\theta dr' 2\pi \theta'
I(\theta') \right/ [\pi \theta^2]},
\end{equation}
where $I(\theta)$ is the azimuthally averaged surface brightness profile
and $\alpha_{\mathrm{lo}}<1$, $\alpha_{\mathrm{hi}}>1$ define the
annulus.  The SDSS has adopted $\alpha_{\mathrm{lo}}=0.8$ and
$\alpha_{\mathrm{hi}}=1.25$.  The Petrosian radius $\petroradius$ is
the radius at which the $\petroratio$ falls below a threshhold value
$\petroratiolim$, set to 0.2 for the SDSS. The Petrosian flux is
defined as the flux within a circular aperture with a radius equal to
$N_P \petroradius$, where $N_P = 2$ for the SDSS.  Petrosian
magnitudes are described in greater detail by \citet{blanton01a} and
\citet{strauss02a}.

Another important measure of flux for galaxies is the SDSS model
magnitude, which is an estimate of the magnitude using the better of a
de Vaucouleurs and an exponential fit to the image. {\tt photo} also
measures the flux in each object using the local PSF is a template,
which is the highest signal-to-noise measurement of flux for point
sources. Finally, {\tt photo} outputs an azimuthally averaged radial
profile for each object.

For the purposes of this paper, we have implemented one more measure
of galaxy magnitude: a \Sersic\ magnitude, following
\citet{sersic68a}.  We fit an axisymmetric \Sersic\ profile to the
azimuthally-averaged radial profile of the galaxy:
\begin{equation}
\label{sersic}
I(r) = A \exp\left[ - (r/r_0)^{1/n} \right],
\end{equation}
where $A$, $r_0$, and $n$ are free parameters quantifying the
amplitude, size, and shape of the surface brightness profile
quantitatively. We do so by convolving $I(r)$ with the double Gaussian
fit to the seeing output by {\tt photo}, and minimizing $\chi^2$ with
respect to the observed radial profile (using the {\tt photo} catalog
output describing the radial profile and its uncertainties, as
described in \citealt{strauss02a}).  In practice, we fit for $r_0$ and
$n$ in the $i$-band and only fit for $A$ in the other bands, fixing
$r_0$ and $n$ to their values in the $i$-band (in analogy to the model
magnitudes output by {\tt photo}). As found by \citet{blanton02a},
galaxies have best fit \Sersic\ indices ranging from around
exponential profiles of $n=1$ (which galaxies tend to be blue, low
luminosity, and low surface brightness) to quite concentrated galaxies
with $n=4$--$5$ (which galaxies tend to be red, high luminosity, and
high surface brightness). A de Vaucouleurs profile, which is generally
ascribed to elliptical galaxies, has $n=4$.

While SDSS Petrosian magnitudes contain over 99\% of the flux within
an exponential profile, they contain only around 80\% of the flux
within a de~Vaucouleurs profile (in the absence of seeing). We can
evaluate the total flux contained in our \Sersic\ model; this estimate
is corrected for seeing {\it and} is an attempt to extrapolate the
profile to infinity. Comparing the SDSS Petrosian magnitudes to the
\Sersic\ magnitudes yields a guess of how much luminosity density we
are missing because of the finite Petrosian aperture size. We have
chosen not to use the SDSS model magnitudes for this purpose because
for the versions of {\tt photo} used for this set of photometry, the
outer radius which {\tt photo} considered in its model fits was too
small to accurately model large objects.

To supply targets selected by the imaging program for the concurrent
spectroscopic program, periodically a ``target chunk'' of imaging data
is processed, calibrated, and has targets selected. These target
chunks never overlap, so that once a set of targets are defined in a
particular region of sky, it never changes in that region. Thus, the
task of determining the selection limits used in any region reduces to
tracking how the target chunks cover the sky, which is done by an
internal SDSS operational database.  Within each target chunk, a
target selection pipeline determines which objects are QSO targets,
galaxy targets, and/or star targets, depending on the properties of
each object. The pipeline selects the Main Sample galaxies used in
this paper according to an algorithm described and tested in
\citet{strauss02a}, which has three most important steps: star/galaxy
separation, the flux limit, and the surface brightness
limit. Expressed quantitatively, these criteria are
\begin{eqnarray}
r_{\mathrm{PSF}} - r_{\mathrm{model}} &>& s_{\mathrm{limit}} \cr
r_{\mathrm{petro}} &<& r_{\mathrm{limit}}, \mathrm{~and}\cr
\mu_{50} &<& \mu_{50,\mathrm{limit}},
\end{eqnarray}
where $r_{\mathrm{petro}}$ is the dereddened Petrosian magnitude in
the $r$ band (using the dust maps of \citealt{schlegel98a}),
$r_{\mathrm{model}}$ is the model magnitude, $r_{\mathrm{PSF}}$ is the
PSF magnitude, and $\mu_{50}$ is the Petrosian half-light surface
brightness of the object in the $r$-band. In practice, the values of
the target selection parameters vary across the survey in a
well-understood way, but for the bulk of the area, they are:
$s_{\mathrm{limit}}=0.3$, $r_{\mathrm{limit}}=17.77$, and
$\mu_{50,\mathrm{limit}}=24.5$. As described in \citet{strauss02a},
there are many more details in galaxy target selection which we do not
have space to discuss here. We note here that objects near the
spectroscopic flux limit are nearly five magnitudes brighter than the
photometric limit; that is, the fluxes are measured at signal-to-noise
of a few hundred.

To drill spectroscopic plates which have fibers on these targets, we
define ``tiling chunks'' which in principle can contain numerous
target chunks (and parts of target chunks). An automatic tiling
pipeline (\citealt{blanton02a}) is then run in order to position tiles
and assign fibers to them, after which plates are designed (that is,
extra fibers are assigned to possibly interesting targets and
calibration fibers are designed) and then drilled. In general, these
tiling chunks {\it will} overlap because we want the chance to assign
fibers to targets which may have been in adjacent, earlier tiling
chunks but were not assigned a fiber. For a target to be covered by a
particular tile, it must be in the same tiling chunk as that tile and
be within 1.49$^\circ$ of the tile center (because the edges of tiles
can extend beyond the tiling chunk boundaries, a particular direction
can be within 1.49$^\circ$ of the tile center but not ``covered'' by
it as defined here). To calculate the survey window function, we then
divide the survey into a number of small regions known as ``sectors''
which are regions which are covered by a unique set of tiles (the same
as the ``overlap regions'' defined in \citealt{blanton01a}). These
sectors are described in {\tt sample10}; the sampling rates are
calculated on a sector-by-sector basis.

The targets are observed using a 640 fiber spectrograph on the same
telescope as the imaging camera. The results of the spectroscopic
observations are treated as follows.  We extract one-dimensional
spectra from the two-dimensional images using a pipeline ({\tt specBS
v4\_8}) created specifically for the SDSS instrumentation (Schlegel et
al. in preparation), which also fits for the redshift of each
spectrum. The official SDSS redshifts are obtained from a different
pipeline (SubbaRao et al. in preparation). The two independent
versions provide a consistency check on the redshift
determination. The results of the two pipelines agree for over 99\% of
the Main Sample galaxies.

\begin{figure}
\figurenum{\fignum}
\plotone{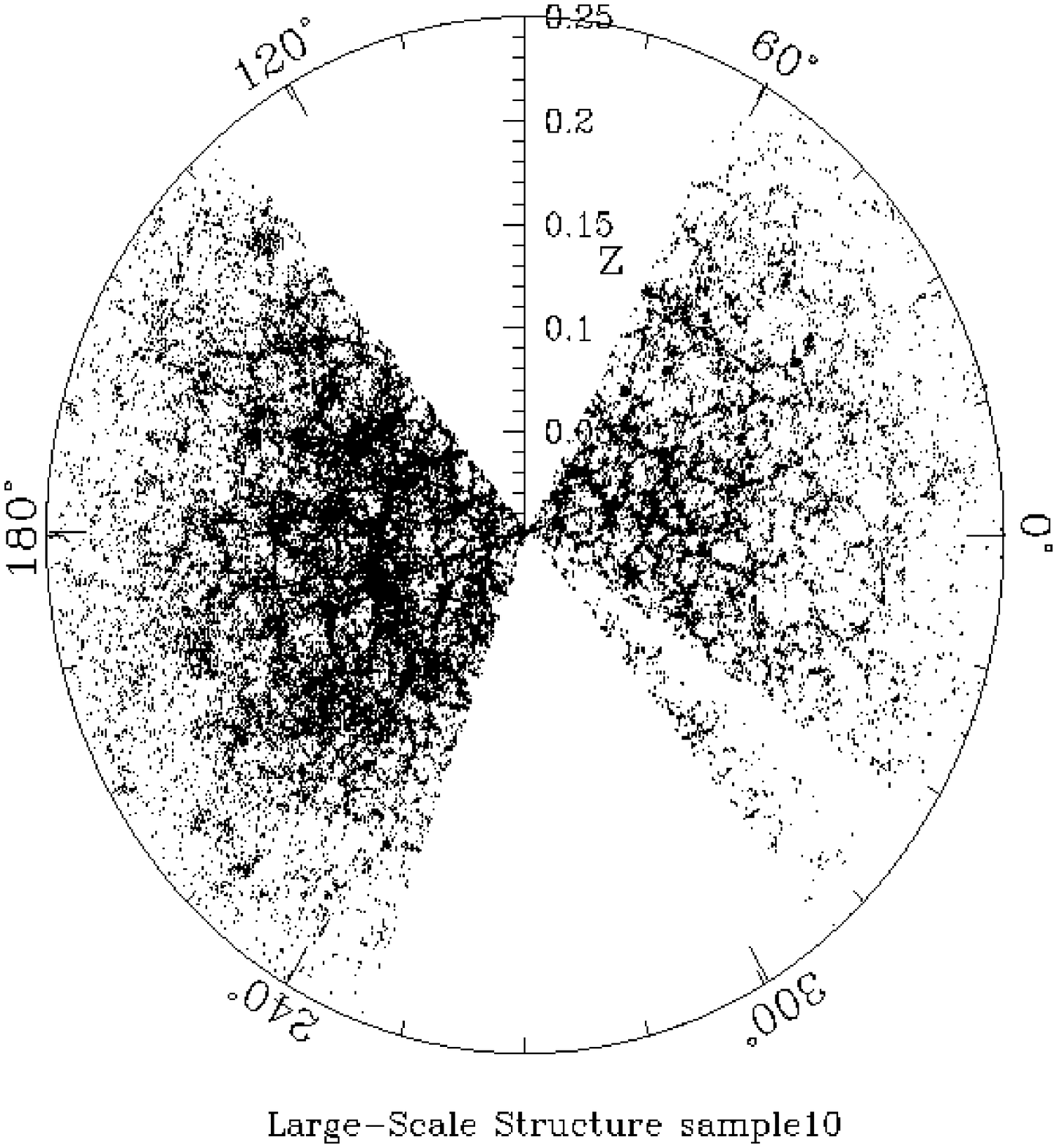}
\caption{\label{pie.sample10} Equatorial distribution of right
ascension and redshift for Main Sample galaxies within 6$^\circ$ of
the Equator in {\tt sample10} of the SDSS. The half of the survey in
the Galactic South (on the right) appears less dense than the Galactic
North (on the left) simply because the imaging near the equator
extends less in the declination direction in the south than in the
north.}
\end{figure}

\begin{figure}
\figurenum{\fignum}
\begin{center}
\resizebox{!}{\textwidth}{\includegraphics{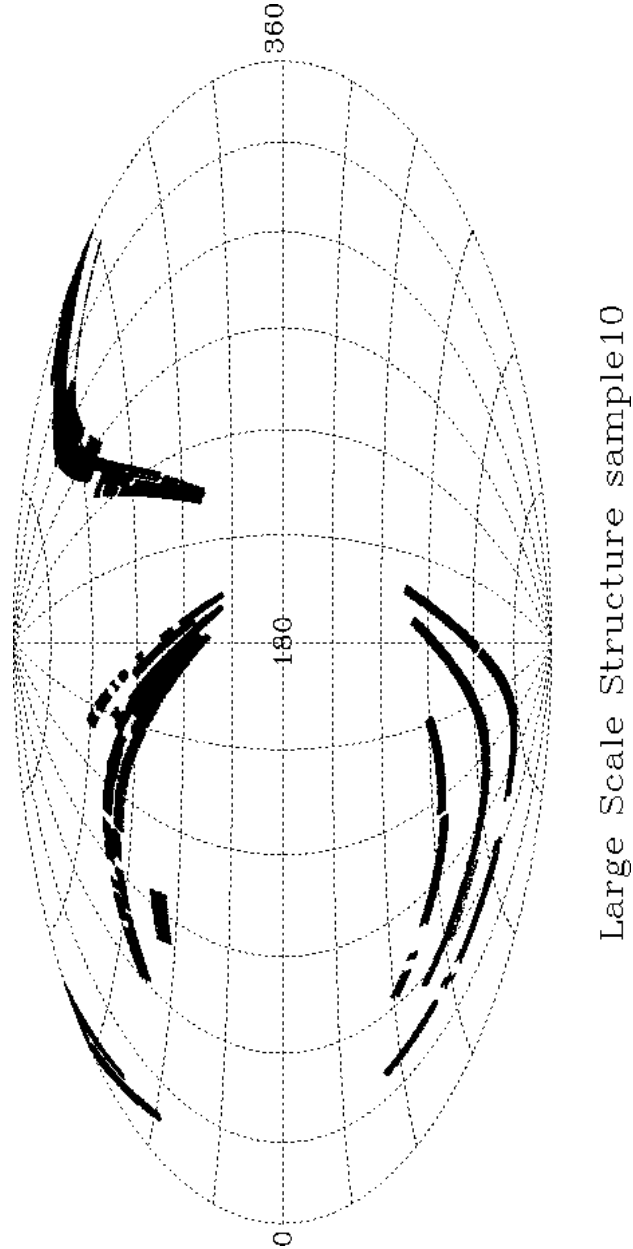}}
\end{center}
\caption{\label{lb.sample10} Distribution in Galactic coordinates of
spectroscopically confirmed galaxies in the SDSS Large-Scale Structure
{\tt sample10}. The effective area covered (the integral of the
SDSS sampling fraction over the covered area) is about 1844 deg$^2$.}
\end{figure}

As of April 2002, the SDSS had imaged and targeted 2,873 deg$^2$ of
sky and taken spectra of approximately 350,000 objects over $\sim
2,000$ deg$^2$ of that area. From these results, we created a
well-defined sample for calculating large-scale structure and galaxy
property statistics, known as Large-Scale Structure {\tt
sample10}. {\tt sample10} consists of all of the photometry for all of
the targets over that area (as extracted from the internal SDSS
operational database), all of the spectroscopic results (as output
from {\tt specBS}), and, most significantly, a description of the
angular window function of the survey and the flux and surface
brightness limits used for galaxies in each area of the sky.  For most
of the area, the same version of the image analysis software used to
create the spectroscopic target list was used in this sample. However,
for the area covered by the Early Data Release (EDR;
\citealt{stoughton02a}) we used the version of the analysis software
used for that data release, since it was substantially better than the
early versions of the software used to target that area. For {\tt
photo}, the most important piece of analysis software run on the data,
the versions used for the photometry range from {\tt v5\_0} to {\tt
v5\_2}. The region covered by this sample is similar to, but not
exactly, the region which will be released in the SDSS Data Release 1
(DR1), scheduled for January 2003 (which will use {\tt photo v5\_3}, a
newer version of the software which among other things improves the
handling of large galaxies).  Figure \ref{pie.sample10} shows the
distribution in right ascension and redshift of Main Sample galaxies
with redshifts in {\tt sample10} within 6$^\circ$ of the
Equator. Figure \ref{lb.sample10} shows the distribution of Main
Sample galaxies with redshifts on the sky in Galactic coordinates.

%Earlier samples have been used to calculate the luminosity function of
%galaxies (\citealt{blanton01a}; {\tt sample5}), to calculate the
%correlation function of galaxies and its dependence on type
%(\citealt{zehavi02a}; {\tt sample7}) and to calculate the void
%distribution and the fractional luminosity density of red galaxies
%(\citealt{hoyle02a}; \citealt{hogg02a}; {\tt sample8}). {\tt sample10}
%is currently being used to provide improved measurement of $\beta$
%from the large-scale redshift distortions of the correlation function,
%to calculate the power spectrum of galaxies, to calculate the
%three-point galaxy correlation function, and to measure the number
%density distribution of broad-band photometric galaxy properties
%(\citealt{blanton02d}).  

\begin{figure}
\figurenum{\fignum}
\plotone{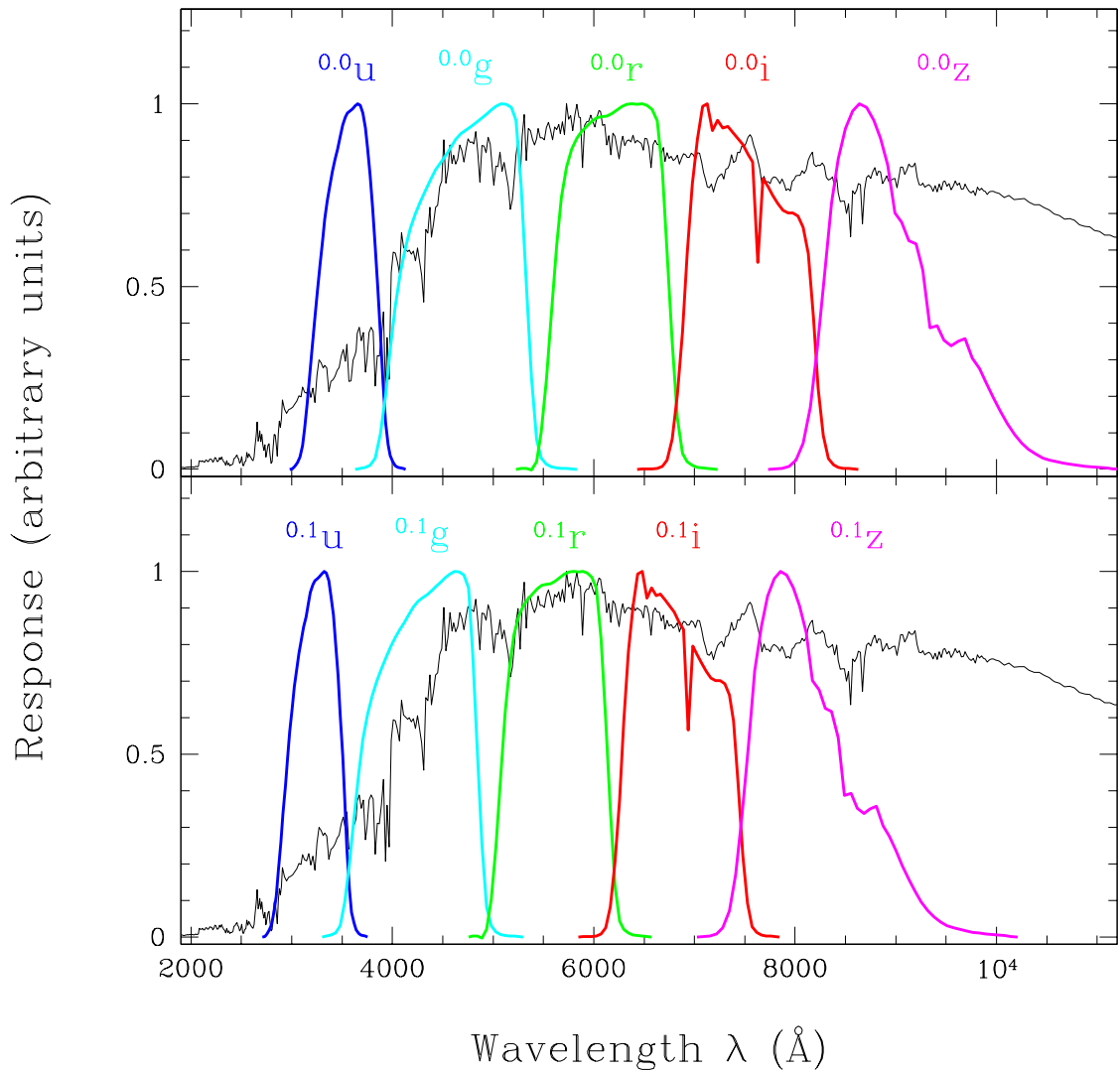}
\caption{\label{response} Demonstration of the differences between the
unshifted SDSS filter system (\band{0.0}{u}, \band{0.0}{g},
\band{0.0}{r}, \band{0.0}{i}, \band{0.0}{z}) in the top panel and the
SDSS filter system shifted by $0.1$ (\band{0.1}{u}, \band{0.1}{g},
\band{0.1}{r}, \band{0.1}{i}, \band{0.1}{z}) in the bottom
panel. Shown for comparison is a 4 Gyr-old instantaneous burst
population from an update of the \citet{bruzual93a} stellar population
synthesis models. The $K$-corrections between the magnitudes of a
galaxy in the unshifted SDSS system observed at redshift $z=0.1$ and
the magnitudes of that galaxy in the $0.1$-shifted SDSS system
observed at redshift $z=0$ are independent of the galaxy's spectral
energy distribution (and for AB magnitudes are equal to $-2.5
\log_{10} (1+0.1)$ for all bands; \citealt{blanton02b}). This
independence on spectral type makes the $0.1$-shifted system a more
appropriate system in which to express SDSS results, for which the
median redshift is near redshift $z=0.1$.}
\end{figure}

We measure galaxy magnitudes through a set of bandpasses which is
obviously constant in the observer frame.  This set of observer frame
magnitudes corresponds to a different set of rest-frame magnitudes
depending on the galaxy redshift.  In order to compare galaxies
observed at different redshifts, we convert all the magnitudes to a
single fixed set of bandpasses.  To perform this conversion, we use
the method of \citet{blanton02b} ({\tt kcorrect v1\_11}). These
routines fit a linear combination of four spectral templates to each
set of five magnitudes, assigning coefficients $a_0$, $a_1$, $a_2$,
and $a_4$. The coefficient $a_0$ to the first template is an estimate
of the flux in the optical range ($3500 \mathrm{\AA} < \lambda < 7500
\mathrm{\AA}$) in ergs s$^{-1}$ cm$^{-2}$; the fractional contribution
of the other coefficients $a_1/a_0$, $a_2/a_0$, and $a_3/a_0$
characterize the spectral energy distribution of the galaxy. The most
significant variation is along $a_3/a_0$.  Taking the sum of the
templates and projecting it onto filter responses, we can calculate
the $K$-corrections from the observed bandpass to any rest-frame
bandpass. In order to minimize the uncertainties in the
$K$-correction, we choose a fixed set of bandpasses blueshifted by
$z=0.1$ in order that they cover the same region of the rest-frame
spectrum as do the observed bandpasses for a galaxy at $z=0.1$ (chosen
because it is near the median redshift of the sample). The bottom
panel Figure \ref{response} shows this set of shifted bandpasses in
the SDSS (\band{0.1}{u}, \band{0.1}{g}, \band{0.1}{r}, \band{0.1}{i},
\band{0.1}{z})); the top panel shows the more commonly used unshifted
restframe bandpasses, measurements of which are more poorly
constrained by our set of data.

\begin{figure}
\figurenum{\fignum}
\epsscale{0.95}
\plotone{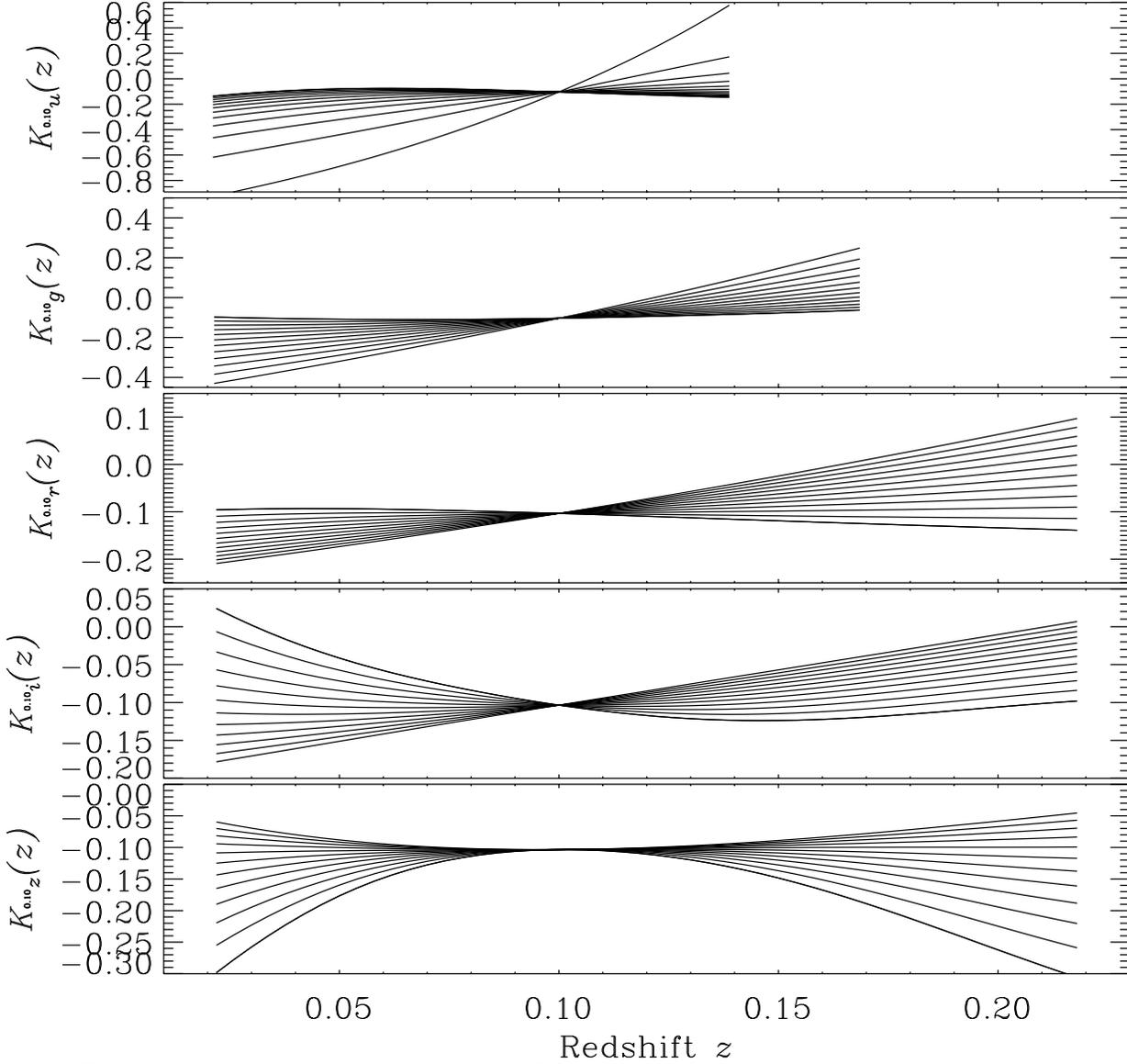}
\epsscale{1.0}
\caption{\label{kcorrect} $K$-corrections for the twelve SED
types used for this paper, for each band. Because we are
$K$-correcting to
bands shifted to $z=0.1$, all galaxies have the same $K$-correction
($-2.5 \log_{10}(1+0.1)$) at $z=0.1$. For this reason, choosing this 
set of bandpasses minimizes our uncertainties in the luminosity
density at $z=0.1$.}
\end{figure}

For this paper, rather than using the full freedom of all four
templates, we instead use $K$-corrections from a restricted set of
models. We fix $a_1/a_0=0$ and $a_2/a_0=0$ (near the mean of the
galaxy distribution in coefficient space). In addition, after fitting
for $a_3/a_0$ we restrict that ratio to be one of the twelve values.
Figure \ref{kcorrect} shows the $K$-corrections in each band as a
function of redshift for each of these twelve values.  The reason to
restrict the $K$-correction this way is to make the likelihood
evaluation more efficient, as described in \citet{blanton02e}. We have
tested the effect of this approximation on our fit for the luminosity
function by taking a smaller set of values (six), finding no
significant differences. Furthermore, if we consider the residuals of
the restricted $K$-corrections in Figure \ref{kcorrect} to the
$K$-corrections found using all four templates, there is no trend with
redshift and the standard deviations of the differences are 0.03 mags
at most (for the $u$ band; the value is much less for the other
bands).

From this sample, we create one subsample for each band, $u$, $g$,
$r$, $i$, and $z$, of galaxies satisfying the apparent magnitude and
redshift limits listed in Table \ref{limits}.  We chose the same
apparent magnitude limits for the bands other than $r$ as
\citet{blanton01a} chose; that paper chose the limits by imposing the
requirement that $<2\%$ of galaxies brighter than the flux limit in
the given band are fainter than $r^\ast=17.6$.  By defining a separate
magnitude-limited sample in each band, we avoid biasing our results in
one band due to the fact that the galaxies were selected in another
band (for example, if we calculated the luminosity function in the $u$
band from a sample limited in the $r$ band, all of the faintest $u$
band galaxies would tend to be red galaxies).  For each band we
include essentially all observed absolute magnitudes in our
samples. However, we of course do not have good constraints at all
absolute magnitudes; for this reason, we consider our model applicable
only to a smaller range of evolution-corrected absolute magnitudes,
which we define from the tenth least luminous object to the most
luminous object. These absolute magnitude limits are also listed in
Table \ref{limits}.

\section{Fitting the Luminosity Function}
\label{method}

In this section, we describe how to recover the number density of
galaxies $\Phi(M,z)$ as a function of absolute magnitude $M$ and
redshift $z$, which we will use to calculate the luminosity density at
$z=0.1$. Each galaxy has a measured magnitude in each band $m$, an
associated uncertainty $\Delta m$, and a redshift $z$. The absolute
magnitude $M_{\band{0.1}{r}}$ may be constructed from the apparent
magnitude $m_r$ and redshift $z$ as follows
\begin{equation}
M_{\mathrm{\band{0.1}{r}}} - 5 \log_{10} h = m_r -
\mathrm{DM}(z,\Omega_0,\Omega_\Lambda, h =1) - K_{\band{0.1}{r}r}(z),
\end{equation}
where as written $\mathrm{DM}(z,\Omega_0,\Omega_\Lambda, h =1)$ is the
distance modulus as determined from the redshift assuming a particular
cosmology (for example, using the formulae compiled by
\citealt{hogg99cosm}) and $h=1$. $K_{\band{0.1}{r}r}(z)$ is the
$K$-correction from the $r$ band of a galaxy at redshift $z$ to the
$\band{0.1}{r}$ band.

In order to fit the distribution of the absolute magnitudes and
redshifts of galaxies, we use a maximum likelihood method which allows
for a generic luminosity function shape (it does not assume a
Schechter function or any other simple form), allows for simple
luminosity and number evolution, and accounts for the estimated
uncertainties in the galaxy fluxes.  The method is described in detail
by \citet{blanton02e}, and we outline it briefly here. It is akin to
the stepwise maximum likelihood (SWML) method of
\citet{efstathiou88a}. However, it includes evolution in our model for
the luminosity function, it accounts for the effects of flux
uncertainties, and (for computational convenience) it uses Gaussian
basis functions rather than top hat basis functions. In addition,
rather than maximizing the likelihood of absolute magnitude given
redshift, it maximizes the joint likelihood of absolute magnitude {\it
and} redshift. This choice makes our estimates more sensitive to
large-scale structure in the sample, and more sensitive to evolution.

Our model for the luminosity-redshift function is
\begin{equation}
\Phi(M,z) = {\bar n} 10^{0.4(z-z_0)P} \sum_k \Phi_k
\frac{1}{\sqrt{2\pi \sigma_M^2}} \exp\left[ -\frac{1}{2} 
\frac{\left(M-M_k+(z-z_0) Q\right)^2}{\sigma_M^2} \right],
%\frac{dV}{dz},
\end{equation}
where the $M_k$ are fixed to be equally spaced in absolute magnitude
and represent the centers of Gaussians of width $\sigma_M$.  $\Phi_k$
are adjustable parameters signifying the amplitudes of the
Gaussians. 

Following \citet{lin99a}, $Q$ represents the evolution in luminosity,
in units of magnitude per unit redshift; $Q>0$ indicates that galaxies
were more luminous in the past. $P$ quantifies the change in the
number density with redshift; we choose this particular
parametrization (again following \citealt{lin99a}) such that $P$
represents the contribution of number density evolution to the
evolution in the luminosity density in units of magnitudes. $P$ can be
interpreted as either due to true evolution in the number density or
due to very large scale structure. Given the size of our data set and
its relative shallowness, we cannot distinguish between these
possibilities; when necessary to, we will interpret $P$ only as
large-scale structure. In any case, our main interest in this paper is
the luminosity density at $z=0.1$, not its evolution, and the
luminosity density is insensitive to reasonable values of $P$.  $z_0$
is the zeropoint redshift, with respect to which we measure the
evolution; for this sample, we choose $z_0=0.1$, the median redshift,
since it is at that redshift that we can observe galaxies in the
luminosity range around $M_\ast$, which contribute the most to the
luminosity density.

In principle we can include large-scale structure in the radial
direction, $\rho(z)$, explicitly in the model, with the constraint
that it have a reasonable power spectrum (since the power in the
density field is constrained mostly by modes which are not purely
radial). However, we have decided not to do so here because it is not
necessary for our goals. 

As described in \citet{blanton02e}, we fit model parameters by
maximizing the likelihood of the model parameters given the data:
\begin{equation}
\prod_i p(Q,P,\ln \Phi_k | M_i, z_i) = \prod_i \frac{p(M_i, z_i |
Q, P, \ln \Phi_k) p(Q,P, \ln\Phi_k)} {p(M_i, z_i)}
\end{equation}
We assume a uniform prior distribution of $\ln\Phi_k$, $Q$, and $P$
(thus guaranteeing that $\Phi_k$ is positive). Because $p(M_i,z_i)$
obviously does not depend on the model parameters, the problem reduces
to minimizing 
\begin{equation}
E = - 2 \sum_i \log p(M_i, z_i | Q, P, \ln \Phi_k). 
\end{equation}
We construct the likelihood $p(M_i, z_i | Q, P, \ln \Phi_k)$ of each
galaxy $i$ by convolving the luminosity function with a Gaussian of
width $\Delta m$ (the estimated apparent magnitude uncertainty defined
above) and constraining the galaxies to satisfy the flux limits of the
survey:
\begin{equation}
p(M, z | Q, P, \ln \Phi_k) = \left\{ \begin{array}{cc}
\Phi(M,z) \otimes G(\Delta m) &
\mathrm{if~} m_{\mathrm{min}} < M + \mathrm{DM}(z) + K(z) - (z-z_0) Q <
m_{\mathrm{max}} \cr 
0 & \mathrm{otherwise}
\end{array} \right.
\end{equation}

The number of parameters required for this fit ($50$--$100$) is small
enough that standard function minimizers can handle the task in a
reasonable amount of time (one hour) on modern workstations
(in our case, a 2 GHz Pentium IV machine), for a sample of $\sim 10^5$
objects. In the fit, we constrain the integral of $\Phi(M,z=0.1)$ to
be unity over our range of absolute magnitude (as listed in Table
\ref{limits} for each band).

The overall normalization ${\bar n}$ cannot be determined from this
likelihood maximization procedure. We use the standard minimum
variance estimator of \citet{davis82a} to perform the normalization. 
\begin{equation}
\label{normalization}
{\bar n} = \frac{\sum_{j=1}^{N_{\mathrm{gals}}} w(z_j)} {\int dV
\phi(z) w(z)},
\end{equation}
where the integral is over the volume covered by the survey between the
minimum and maximum redshifts used for our estimate.  The weight for
each galaxy is
\begin{equation}
w(z)=\frac{f_t}{1+ {\bar n} 10^{0.4 P (z-z_0)} J_3 \phi(z)},
\end{equation}
and the selection function is
\begin{equation}
\label{selfunc}
\phi(z) = \frac{{\int_{L_{\mathrm{min}}(z)}^{L_{\mathrm{max}}(z)} dL\,
\Phi(L,z) }}{
{\int_{L_{\mathrm{min}}}^{L_{\mathrm{max}}} dL\,
\Phi(L,z) }
},
\end{equation}
where $f_t$ is the galaxy sampling rate determined at each position of
sky as the fraction of targets in each sector that were successfully
assigned a classification. The integral of the correlation function
is:
\begin{equation}
J_3 = \int_0^\infty dr\, r^2 \xi(r) = 10000~h^{-3}~\mathrm{Mpc}^3.
\end{equation}
Clearly, because ${\bar n}$ appears in the weight $w(z)$, it must be
determined iteratively, which we do using the simple estimator ${\bar n}
= (1/V) \sum 1/\phi(z_j)$ as an initial guess. 

To determine the uncertainties in our fit, we use thirty jackknife
resamplings of the data. In each sampling, we omit $1/30$ of the
effective area of sky (meaning, the area weighted by the sampling rate
$f_t$). Each omitted area is a nearly contiguous set of sectors. This
jackknife resampling procedure thus includes, to the extent possible,
the uncertainties due to large-scale structure and calibration errors
across the survey.  Effectively, it includes the effects of errors
which are correlated with angular position on the largest scales.
Taking the results of all thirty fits to the data, we calculate the
covariance between all of our measured parameters using the standard
formula
\begin{equation}
\Avg{\Delta x_i \Delta x_j} = 
\frac{N-1}{N} \sum_{i} (x_i - {\bar x}_i) (x_j - {\bar x}_j)
\end{equation}
The uncertainty correlation matrix is then defined in the standard
way: $r_{ij} = \avg{\Delta x_i \Delta x_j}/(\avg{\Delta x_i^2}
\avg{\Delta x_j^2})^{1/2}$. 

We are not interested in the uncertainty correlations between the amplitudes of
each Gaussian, because obviously neighboring Gaussians will be highly
covariant. For this paper, we will not even be interested in the
covariances between overall amplitude at different luminosities of the
luminosity function calculated from the sum of the Gaussians. However,
we will list in tables the covariances between the luminosity density,
the evolution parameters, and overall measures of the shape, such as
$M_\ast$ and $\alpha$ for the best fit Schechter function of each
luminosity function. It is also important to track the covariances
among the luminosity densities in all of the bands; because
large-scale structure is an important source of uncertainty, the
luminosity densities are highly covariant, and ignoring this
covariance would lead to overconfidence in any fit to the stellar
density in galaxies based on this data (and {\it underconfidence} in 
our knowledge of the relative luminosity density in different bands).

Note that the true uncertainties in the luminosity density may be
dominated by the uncertainties in the overall photometric calibration
or by the fraction of flux contained within the Petrosian aperture for
the galaxies that contribute to the luminosity density, while the
uncertainties in the level of evolution recovered may be dominated by
possible systematic errors in the $K$-corrections, as well as a
systematic dependence of the fraction of light contained within a
Petrosian magnitude as a function of redshift (due to the effects of
seeing).

\section{Results}
\label{results}

We have applied the procedure described in the Section \ref{method} to
the data described in Section \ref{data}. Our results are summarized
in this section.

\subsection{Luminosity Functions}

\begin{figure}
\figurenum{\fignum}
\plotone{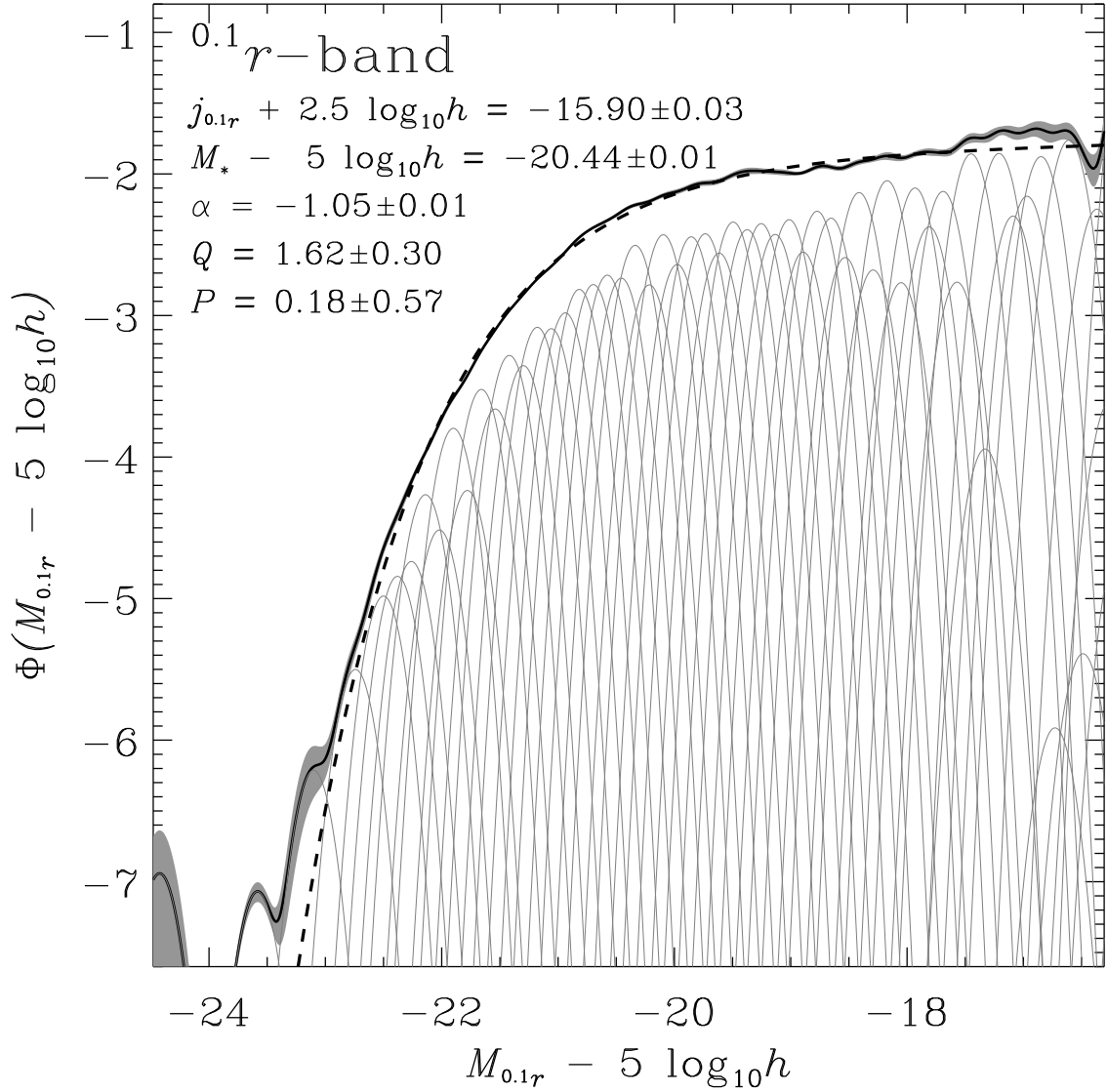}
\caption{\label{lfr} Luminosity function in the \band{0.1}{r}
band. The thick solid line is the luminosity function fit; the thin
solid lines are the individual gaussians of which it is composed. The
grey region around the luminosity function fit represents the
1$\sigma$ uncertainties around the line; naturally, these
uncertainties are highly correlated with each other. The dashed line
is the Schechter function fit to the result.  The luminosity density,
the evolution parameters, and the parameters of the Schechter function
are listed in the figure.}
\end{figure}

\begin{figure}
\figurenum{\fignum}
\plotone{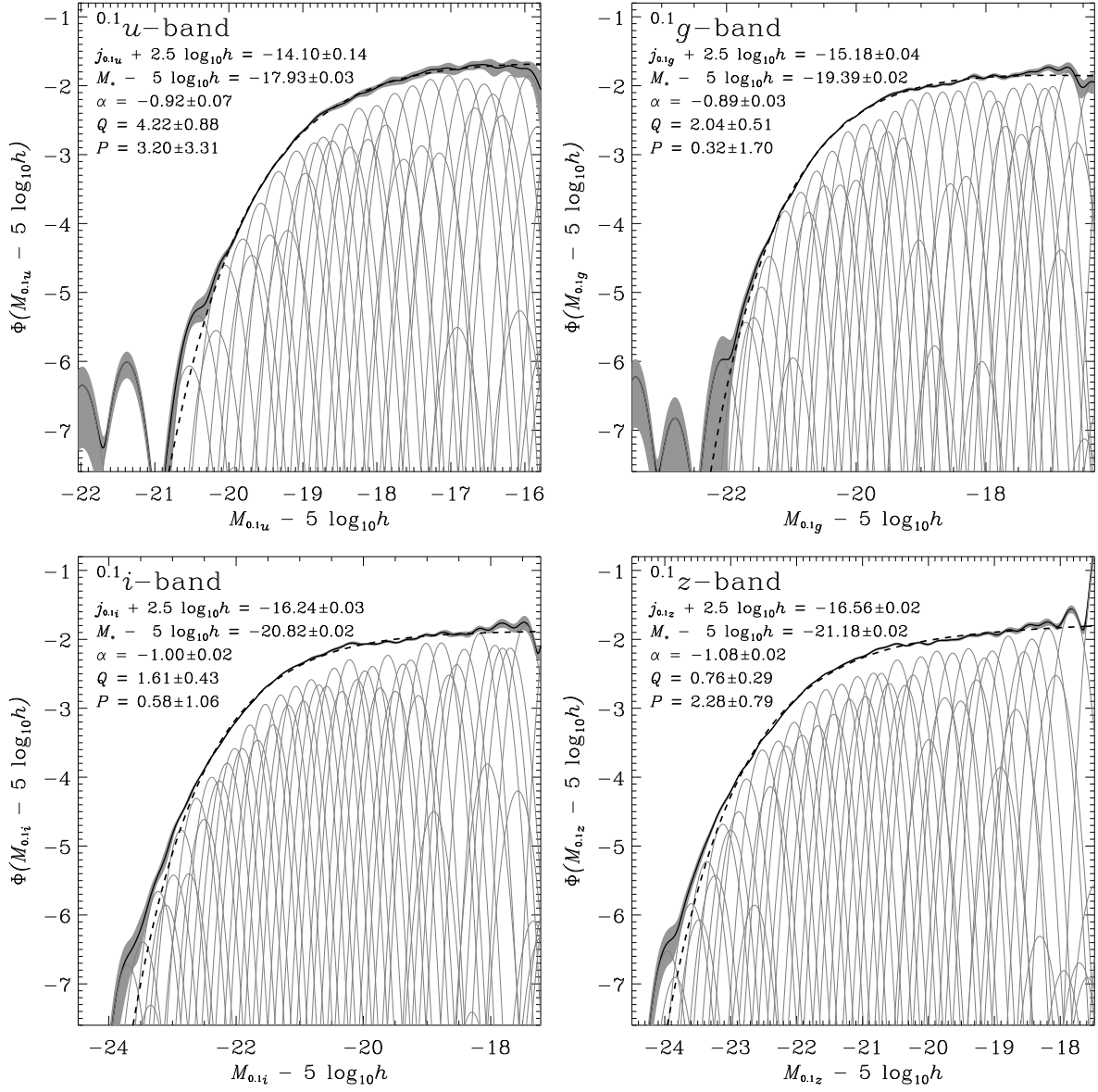}
\caption{\label{lfugiz} Same as Figure
\ref{lfr}, for the \band{0.1}{u}, \band{0.1}{g}, \band{0.1}{i}, and
\band{0.1}{z} bands.}
\end{figure}

Figures \ref{lfr} and \ref{lfugiz} show the galaxy luminosity function
in the $\band{0.1}{u}$, $\band{0.1}{g}$, $\band{0.1}{r}$,
$\band{0.1}{i}$, and $\band{0.1}{z}$ bands, assuming $\Omega_0=0.3$
and $\Omega_\Lambda=0.7$. The thick black line shows our best fit
luminosity function. The thin black lines show the Gaussians which sum
to form the full luminosity function. The grey region surrounding the
thick black line indicates the 1$\sigma$ uncertainties in the
luminosity function --- of course, these uncertainties are all
correlated with one another, and are closer to representing the
uncertainties in the overall normalization of the function than the
individual uncertainties at each magnitude. The best-fit $Q$ and $P$
evolution parameters are listed in the figure.

We have taken the thick black lines and their uncertainties and fit a
Schechter function to each curve. The dotted lines in Figures
\ref{lfr} and \ref{lfugiz} represent the best fit Schechter functions,
which provide a reasonable fit to our non-parametric results. The
luminosity density we list in the figure, expressed as the absolute
magnitude from galaxies in an $h^{-3}$ Mpc$^3$ on average, is the
result of integrating this Schechter function fit over all
luminosities. The values associated with the Schechter function are
listed in Table \ref{schtable}. We list results for the
($\Omega_0=1.0$, $\Omega_\Lambda=0.0$) and ($\Omega_0=0.3$,
$\Omega_\Lambda=0.0$) cosmologies as well. We have found that to an
accuracy of about 3\%, we can convert the results of one cosmology to
those of another by scaling $\phi_\ast$ by the inverse ratio of the
comoving volumes at $z=0.1$ between the two cosmologies, and by
scaling $M_\ast$ by the difference of the distance moduli at $z=0.1$
for the two cosmologies. We therefore recommend this procedure for
readers interested in comparing our results to those in some other
cosmological model.

Table \ref{lftable} lists some salient quantitative measurements of
the luminosity function in each band, including the evolution
parameters and the luminosity density (expressed in magnitudes, solar
luminosities, and flux at the effective filter wavelength) for a
Mpc$^3$. To obtain the physical expressions of the luminosity density,
we used measurements of the SDSS camera response performed by Mamoru
Doi, which James Gunn combined with estimates of the atmospheric
extinction as a function of wavelength at $1.3$ airmasses (to which
all SDSS observations are calibrated) and the primary and secondary
mirror reflectivities. Projecting the solar model of \citet{kurucz91a}
onto these bandpasses (shifted to $z=0.1$) yields the absolute solar
AB magnitudes:
\begin{equation}
M_{\odot,\band{0.1}{u}} = 6.80\mathrm{;} \quad
M_{\odot,\band{0.1}{g}} = 5.45\mathrm{;} \quad
M_{\odot,\band{0.1}{r}} = 4.76\mathrm{;} \quad
M_{\odot,\band{0.1}{i}} = 4.58\mathrm{;} \quad
M_{\odot,\band{0.1}{z}} = 4.51
\end{equation}
The luminosity densities expressed in ergs s$^{-1}$ \AA$^{-1}$ are
calculated from the AB magnitudes as follows. First, we use the
equation which relates an AB magnitude to the effective flux
density at the effective wavelength,
\begin{equation}
\label{fluxeq}
f_\lambda = 
(3.631 \times 10^{-20}
\mathrm{ergs~cm}^{-2}\mathrm{~s}^{-1}\mathrm{~Hz}^{-1})  
\frac{c}{\lambda_{\mathrm{eff}}^2}
10^{-0.4 m},
\end{equation}
to convert the absolute magnitudes in an $h^{-3}$ Mpc$^3$ to the flux
density which would be observed if an average $h^{-3}$ Mpc$^3$ of the
universe were compressed to a point source and placed 10 pc distant
from the observer.  Second, we multiply this value by $4\pi
(10\mathrm{~pc})^2$ to obtain the average luminosity per unit
wavelength at the effective wavelength in an $h^{-3}$ Mpc$^3$.  The
``effective wavelength'' of a passband with a quantum efficiency
$R(\lambda)$ is defined:
\begin{equation}
\label{lden}
\lambda_{\mathrm{eff}} = \exp\left[ 
\frac{\int d(\ln\lambda) R(\lambda) \ln\lambda}
{\int d(\ln\lambda) R(\lambda)} \right],
\end{equation}
following \citet{fukugita96a} and \citet{schneider83a}. The effective
flux density defined above is that which an AB standard source
($f_\lambda(\lambda) \propto \lambda^{-2}$) of magnitude $m$ in
passband $R$ would have at the effective wavelength. Both of these
quantities are obviously only nominal since, in any case, the average
spectrum of galaxies is nothing like an AB standard source, but it
does give a sense of the physical flux associated with a magnitude. We
also list $f_{\mathrm{np}}$, the fraction of the integrated luminosity
density of the Schechter luminosity function included in the
non-parametric estimate of the luminosity density. Results for the
cosmologies ($\Omega_m = 0.3$, $\Omega_\Lambda = 0.0$) and ($\Omega_m
= 1.0$, $\Omega_\Lambda = 0.0$) are also listed.

It is worth asking how well this model reproduces the number counts of
galaxies as a function of redshift and absolute magnitude. Figure
\ref{zhist} shows the redshift distribution of galaxies in our sample
for quartiles in absolute magnitude in the \band{0.1}{r} band as a
thick histogram. The expectation from our model fit (based on Monte
Carlo realizations of the sample) is shown as the thin histogram.
Figures \ref{uqa}--\ref{zqa} show the counts in bins of absolute
magnitude for several slices in redshift. Again, the thick histogram
represents the data and the thin histogram represents the model.  The
model appears to the eye to be a reasonable fit to the data. However,
it is clear that there are statistically significant discrepancies in
these figures (note the large number of objects in each bin). Some of
these discrepancies occur because there is large-scale structure in
the sample; however, it is possible that we will need to introduce a
more sophisticated model for the evolution of the galaxies or for the
relationship between density and luminosity in order to explain all of
these discrepancies. As discussed in the conclusions, we postpone the
investigation of these issues to a later paper.

%Figure \ref{ugizlf} shows the galaxy luminosity functions in the
%$\band{0.1}{u}$, $\band{0.1}{g}$, $\band{0.1}{i}$, and $\band{0.1}{z}$
%bands, in the same style as Figure \ref{rlf}. Again, the Schechter
%function is a reasonable fit. Figure \ref{ugizqa} shows the number
%counts as a function of redshift and absolute magnitude for each
%band. Again, the model is an decent representation of the data.

%Table \ref{lumdens} lists the properties of the fits to all bands,
%including the luminosity density at $z=0.1$, expressed both in
%absolute magnitudes in a Mpc$^3$ and in ergs/s/Angstrom/Mpc$^3$, the
%evolution parameters $P$ and $Q$, the Schechter model parameters, and
%the fraction of the luminosity density contained in our absolute
%magnitude range compared to the luminosity density in the Schechter
%model extrapolated to zero luminosity galaxies. We list the results
%assuming three cosmologies: ($\Omega_0=0.3$, $\Omega_\Lambda=0.7$),
%($\Omega_0=0.3$, $\Omega_\Lambda=0.0$), and ($\Omega_0=1.0$,
%$\Omega_\Lambda=0.0$).

As described above, we have quantified our uncertainties by taking 30
jacknife resamplings of the data. In Tables
\ref{ucorrtable}--\ref{zcorrtable} we display the resulting
correlation matrices between various properties of our fit for each
band. Note that many of the parameters are highly correlated. In
particular we note that $M_\ast$ and $\alpha$, which characterize the
shape of the Schechter luminosity function, are highly correlated. One
should be cautious when quoting $M_\ast$ values as representative of
the ``typical luminosity'' of a luminous galaxy. This statement is
true to an extent, but not in detail.

Table \ref{fulljk} displays the correlation matrix between the
luminosity densities and evolution parameters $Q$ in all of the
bands. Note that this matrix is entirely positive. In any particular
resampling, if the luminosity density is slightly high in one band
relative to the mean, all the bands are slightly high; when one band
evolves a bit more strongly, so do they all. This correlation occurs
because our errors are dominated by large-scale structure, which
affects all the bands simultaneously. These correlations are strong,
so it is important to account for this correlation matrix of
uncertainties when using these results.

The top panel of Figure \ref{mdenq} shows the luminosity density
as a function of wavelength at $z=0.1$ using the results of all five
bands. 
%For comparison, the spectrum of an 8 Gyr old instantaneous
%burst of solar metallicity (using the models of \citealt{bruzual93a}) 
%is shown.

\subsection{Galaxy Luminosity Density Evolution} 

% SERSIC RESULTS (NP)
% MDENFULL        DOUBLE    =       -13.95
% MDENFULL        DOUBLE    =       -15.21
% MDENFULL        DOUBLE    =       -15.94
% MDENFULL        DOUBLE    =       -16.30
% MDENFULL        DOUBLE    =       -16.67

% Diffs to add to get \Sersic: [-0.01,-0.03,-0.03,-0.06,-0.10]

One product of our fit is $Q$, the evolution of the luminosity
density. However, we caution that we have restricted our model such
that galaxies of all luminosities evolve identically. This assumption
is probably incorrect, because different galaxy types have different
luminosity function and are expected to evolve in different ways
(since their differing colors obviously imply different star-formation
histories). Furthermore, our understanding of our photometric error
model, particularly in $u$ and $z$, is currently rather
primitive. While we believe that the flaws in our model for the
evolution do not greatly bias the main result of this paper, the
luminosity function and luminosity density at $z=0.1$, we do warn the
reader that because of the deficiencies of our model the measured $Q$
values might be biased.

We show our results for the luminosity evolution, $Q$, and its
uncertainties, in the bottom panel of Figure \ref{mdenq}, for all the
bands. The \band{0.1}{u} band has very strong evolution ($Q \sim 4$),
although with large uncertainties. The other bands all have $Q \sim
1$--$2$. The evolution in these bands is generally consistent with the
evolution of a relatively old stellar population.
%For example, we show for reference the expected rate of
%evolution for the 8 Gyr instantaneous burst population in terms of
%$Q$; 
%the observed evolution is not far higher than expected for this
%simple model.  
However, we will consider fits to theoretical models
more carefully in a separate paper, and draw no particular conclusions
here.

It is likely that these evolution results reflect the evolution of the
more luminous galaxies in the sample, for two reasons. First, more
luminous galaxies are observable over a larger redshift range. Second,
the low luminosity end of the luminosity function is nearly a
power-law, which makes it more difficult to detect evolution. If $Q$
traces the evolution of primarily the luminous galaxies, it might
explain why the evolution in the \band{0.1}{u} band is unusually
large; the most luminous objects in the \band{0.1}{u} band are the
blue, exponential profile objects, which we expect to evolve more
rapidly than red, concentrated galaxies.

For the flat, $\Lambda$-dominated cosmology, $\band{0.1}{u}$ and
\band{0.1}{z} bands have $P\ne 0$, though at low significance in the
case of $\band{0.1}{u}$. We do not believe that $P$ reflects true
number density evolution, which we regard as {\it a priori} unlikely,
especially since the other bands have $P$ consistent with
zero. Instead $P$ probably is compensating for inaccuracies in some
other aspect of our model. In the case of the $\band{0.1}{u}$ band it
may compensate for the fact that our model does not include
large-scale structure. In the case of the $\band{0.1}{z}$ band, our
error model could be insufficient, or the $K$-corrections might be
incorrect (though they are small in any case).  Thus, one {\it might}
claim that the luminosity density evolution we measure should be
$Q+P$; however, we think it better to interpret a non-zero $P$ to
reflect large-scale structure in the sample rather than true
evolution. A more sophisticated approach would be to treat the problem
in more detail by fitting for the radial large-scale structure
simultaneously with the luminosity function.

We note in passing that the differences in $P$ and $Q$ between the
flat, $\Lambda$-dominated cosmology and the other cosmologies are
consistent with the differences in the redshift dependence of the
luminosity distance and the differential comoving volume among these
cosmologies. We do not regard the low value of $P$ for $g$, $r$, and
$i$ in the flat, $\Lambda$-dominated cosmology necessarily as evidence
that that cosmological model is correct.

\section{Systematics Tests of Petrosian Magnitudes}
\label{systematics}

One of the challenges in studying the luminosity function and its
redshift dependence is that one must verify that one's measurements of
galaxy luminosity are consistent as a function of redshift. That is,
one must test whether observing an identical galaxy at different
redshifts will yield the same fraction of the total independent of
redshift. Otherwise, systematic errors in the measurement of galaxy
luminosities as a function of redshift could masquerade as evolution. 

Several :w effects can yield artificial redshift dependence in one's
determination of galaxy flux (which is to say, redshift dependence
which cannot be accounted for using the distance modulus and the
$K$-correction). First, the increased physical size corresponding to
the angular PSF width for galaxies at high redshift can change both
the metric size (in physical units at the galaxy) of the aperture used
to calculate the galaxy flux {\it and} the amount of light scattered
outside of any particular aperture. Second, color gradients in
galaxies mean that the different radii in the galaxy profile
$K$-correct differently. Since in the SDSS the size of the aperture is
based on the shape of the profile, this effect can change the metric
size of the aperture used to calculate the galaxy flux. Third,
cosmological surface brightness dimming can make the outermost
measurable isophote have a smaller metric size.

Because the SDSS has high signal-to-noise ratio imaging and because it
uses Petrosian rather than isophotal magnitudes, the third effect is
negligible. However, Petrosian magnitudes are sensitive to some degree
to seeing; as an object becomes small with respect to the seeing
width, the fraction of the total flux measured by the Petrosian
magnitudes tends to that fraction for a pure PSF ({\it e.g.}
\citealt{blanton01a}, \citealt{strauss02a}). For the SDSS, the
fraction of the total light in a pure PSF measured by the Petrosian
magnitude is about 95\%. Thus, galaxies with exponential profiles will
tend to have their measured fluxes reduced in the presence of bad
seeing from the fraction of nearly 100\% for the no-seeing case. On
the other hand, galaxies with de Vaucouleurs profiles will tend to
have their measured fluxes increased from the fraction of only 82\%
for the no seeing case. Furthermore, the apertures of the Petrosian
magnitudes are determined by the $r$-band image of the galaxy, which,
while it is less sensitive to differential $K$-corrections than bluer
bands, is still somewhat dependent on these effects.

Several methods exist for estimating what fraction of the flux is
outside the SDSS Petrosian apertures. First, one could compare SDSS
imaging to deeper imaging. Second, one could stack many SDSS images of
different galaxies but of similar types, and use the composite profile
to characterize the fraction of light missing outside of a Petrosian
aperture. Third, one can use a reasonable model for galaxy light
profiles to extrapolate to a ``total'' flux for each object. Here, we
take the third approach, using the \Sersic\ magnitudes described in
Section \ref{data} as a seeing-corrected, ``total'' magnitude. In
addition, because the \Sersic\ fit is based on the $i$-band image, the
issue of the $K$-correction of the galaxy profile is negligible (as the
$K$-corrections in the $i$-band are small) for the redshifts
considered in this paper.

Figure \ref{check_sersic} shows the differences between \Sersic\ and
Petrosian magnitudes in each band, as a function of \Sersic\ index $n$
and redshift $z$. Clearly, the difference is sensitive to the \Sersic\
index and is larger for galaxies which are similar to de Vaucouleurs
galaxies, in accordance with the estimates in \citet{blanton01a} of
the fraction of flux included in the SDSS Petrosian magnitudes for
different profile types. For $u$ and $g$ the \Sersic\ magnitude is
significantly fainter than the Petrosian magnitude at low \Sersic\
index. This fact {\it might} reflect color gradients within
exponential galaxies. The shape of the \Sersic\ model we use is
determined in the $i$-band; if the effective size of the galaxy is
larger in bluer bands, known to be the case for spiral galaxies, the
flux determined by fitting the \Sersic\ model amplitude will be
smaller than the flux determined by counting the flux within an
unweighted aperture.  The right panel shows a linear regression versus
redshift. The mean offset at $z=0.1$ ranges from 0.14 for the
$\band{0.1}{u}$ band to -0.14 for the $\band{0.1}{z}$ band. Again,
this result suggests that the \Sersic\ magnitude traces a redder
population, perhaps as a consequence of color gradients in galaxies.

For our purposes here, we are interested primarily in the differences
in the resulting luminosity density from \Sersic\ magnitudes relative
to Petrosian magnitudes.  When we fit luminosity functions to all five
bands using the \Sersic\ magnitudes, we obtain luminosity densities
offset from the Petrosian values as follows:
\begin{eqnarray} 
\label{sersicdiff}
j_{\band{0.1}{u}}(\mathrm{Sersic}) &=&
j_{\band{0.1}{u}}(\mathrm{Petrosian }) -0.01\cr
j_{\band{0.1}{g}}(\mathrm{Sersic}) &=&
j_{\band{0.1}{g}}(\mathrm{Petrosian }) -0.03\cr
j_{\band{0.1}{r}}(\mathrm{Sersic}) &=&
j_{\band{0.1}{r}}(\mathrm{Petrosian }) -0.03\cr
j_{\band{0.1}{i}}(\mathrm{Sersic}) &=&
j_{\band{0.1}{i}}(\mathrm{Petrosian }) -0.06\cr
j_{\band{0.1}{z}}(\mathrm{Sersic}) &=&
j_{\band{0.1}{z}}(\mathrm{Petrosian }) -0.10
\end{eqnarray} 
The values of $Q$ for \Sersic\ magnitudes are consistent with the
results for Petrosian magnitudes. We caution that the estimates of the
luminosity density using \Sersic\ magnitudes are not necessarily more
correct than the estimates from Petrosian magnitudes; because the
\Sersic\ profile is not a perfect fit to the observed radial profiles of
galaxies, there is likely to be a bias in the resulting estimates of
luminosity density associated with \Sersic\ magnitudes. However, the
small differences in Equation (\ref{sersicdiff}) suggest that the
differences are also small between either of these estimates and the
luminosity density we would derive using true ``total'' magnitudes.

\section{Comparison with Other Results}
\label{comparison}

\subsection{Galaxy Luminosity Density in Other Bands}

In order to compare our results to those of other investigators, we
have used the routines in {\tt kcorrect v1\_14} (a slight update on
the version {\tt kcorrect v1\_11} used elsewhere in this paper and in
\citealt{blanton02b}) to fit the SED of the universe and derive the
luminosity densities in other bands. In particular, we list in Table
\ref{kuniv} results for the unshifted SDSS bandpasses for comparison
to \citet{blanton01a}, for the unshifted $B$, $V$, $R$ and $I$ Johnson
bandpasses (using the responses listed by \citealt{bessell90a}), and
for $b_j$ for comparison to 2dFGRS. To create the $b_j$ result we
applied the equation $b_j - B = -0.28 (B-V)$ to the results in the
Johnson bandpasses, originally from \citet{blair82a}. We derive these
luminosity densities both at $z=0.1$ and (by using $Q$ to evolve the
SDSS results) $z=0$. For comparison, we show in the same table the
results of \citet{blanton01a} and \citet{norberg02a}. For the
\citet{norberg02a} we use the mean evolutionary correction from their
Figure 8 to evaluate a luminosity density at $z=0.1$. There is only
0.09 magnitude difference between our result in $b_j$ at $z=0.1$ and
theirs; this difference is less than 1$\sigma$ taking into account the
2dFGRS statistical uncertainties.

The results of Table \ref{kuniv} reflect a general agreement between
these different determinations of the luminosity function.  In
contrast to these results, \citet{blanton01a} reported a higher
luminosity density than found in other surveys, such as the 2dFGRS
results of \citet{folkes99a} and the Las Campanas Redshift Survey
(LCRS; \citealt{shectman96a}) results of \citet{lin96a}. Most
recently, \citet{liske02a} found that the $\band{0.0}{g}$-band
luminosity function of \citet{blanton01a} overpredicted the number
counts of galaxies found in the Millenium Galaxy Catalog. Furthermore,
as pointed out by \citet{wright01a}, the luminosity density in the
$\band{0.0}{z}$-band reported by \citet{blanton01a} was unreasonably
high relative to the luminosity density found by \citet{cole00a} in
$\band{0.0}{J}$ and $\band{0.0}{K}$. While our luminosity density is
still considerably higher than that found by the LCRS, we no longer
disagree significantly with 2dFGRS.  Naturally, the question arises as
to why the discrepancy existed in the first place.

\begin{figure}
\figurenum{\fignum}
\plotone{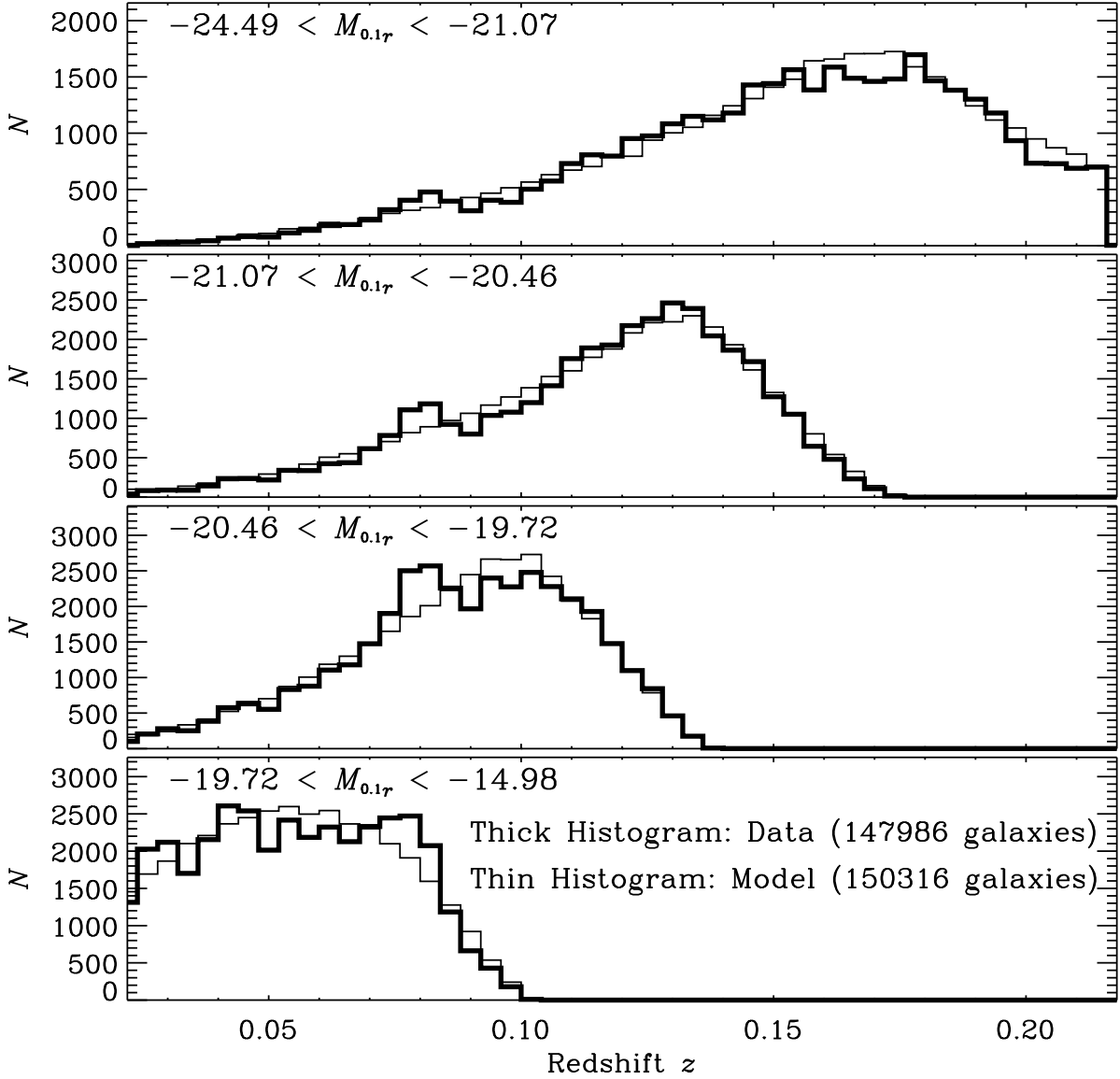}
\caption{\label{zhist} Redshift distribution of the \band{0.1}{r} band
sample, for each quartile (weighted by number) in absolute magnitude.
The thick line represents the data; the thin line is a Monte Carlo
representation of the model, including the selection effects in the
survey. In this figure and in the following Figures
\ref{uqa}--\ref{zqa}, the model is a decent representation of the
data, but not a perfect one; much of the difference is likely to be
due to large-scale structure, but it is possible that further
complications of our evolution model or our error model might be
necessary to fully reproduce the data.}
\end{figure}

\begin{figure}
\figurenum{\fignum}
\plotone{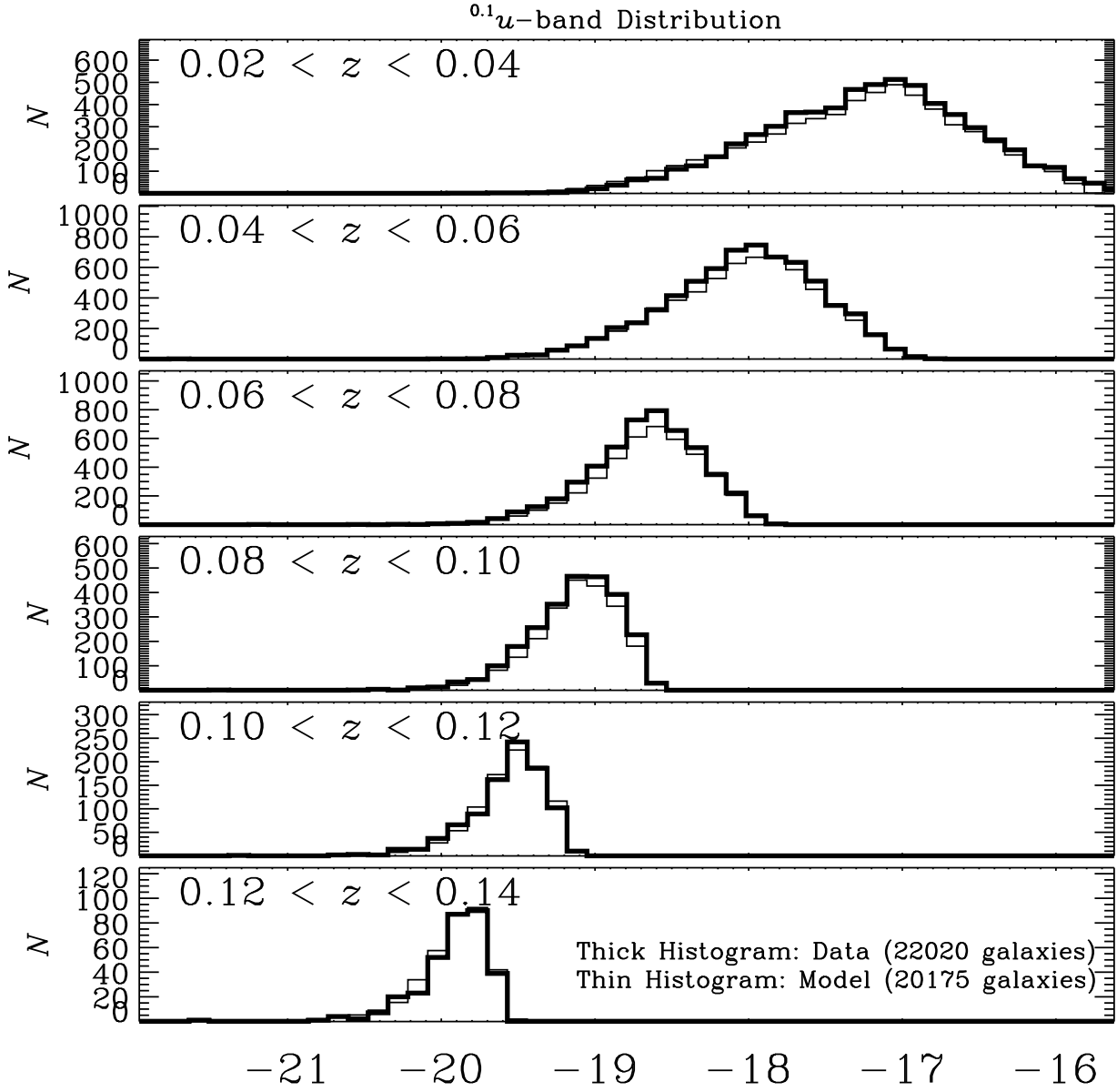}
\caption{\label{uqa} Absolute magnitude 
distribution in the
\band{0.1}{u} band, for several slices of redshift. The thick line 
reperesents the data; the thin line is a Monte Carlo representation of
the model, including the selection effects in the survey. }
\end{figure}

\begin{figure}
\figurenum{\fignum}
\plotone{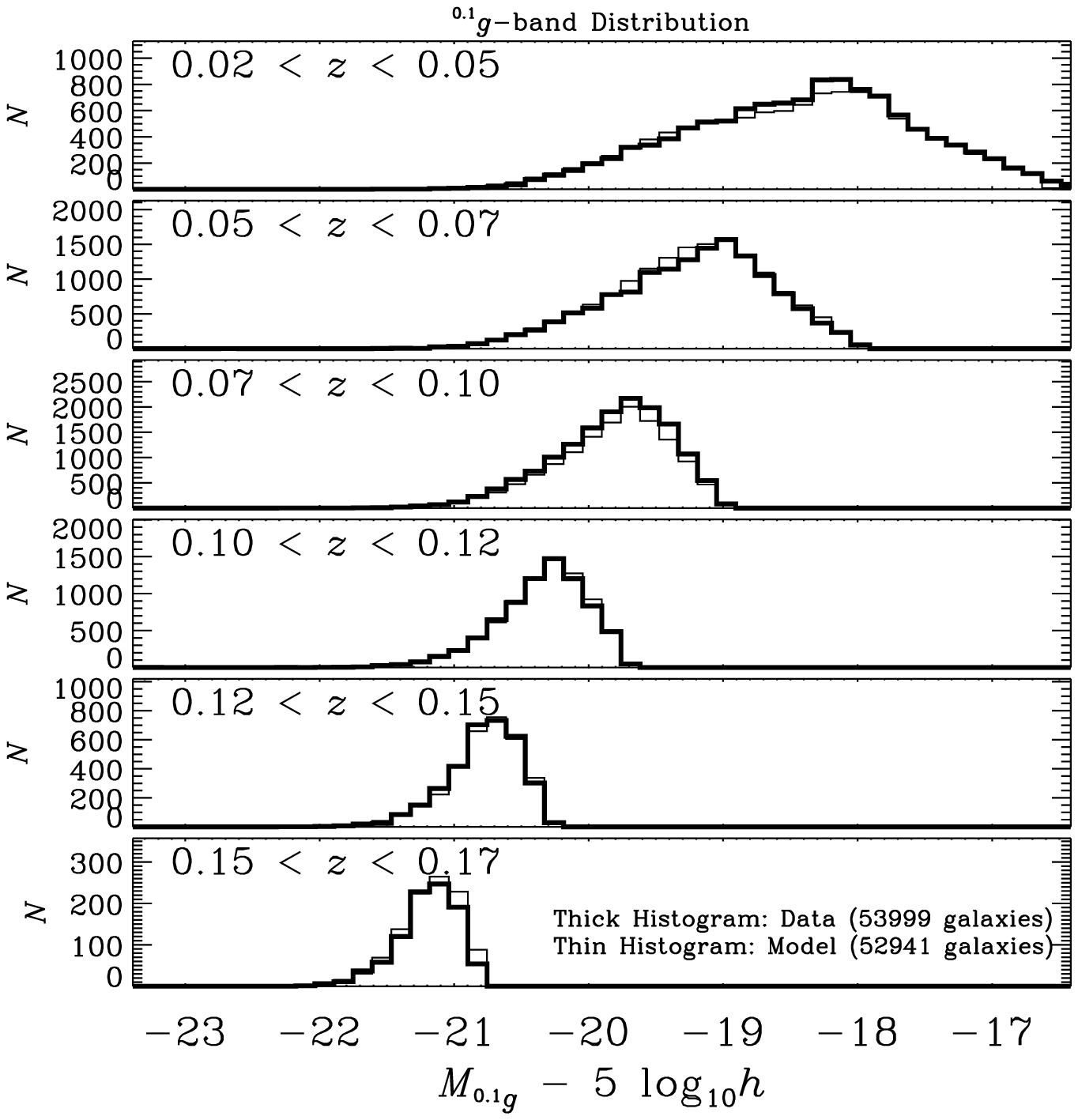}
\caption{\label{gqa} Same as Figure
\ref{uqa}, for the \band{0.1}{g} band.}
\end{figure}

\begin{figure}
\figurenum{\fignum}
\plotone{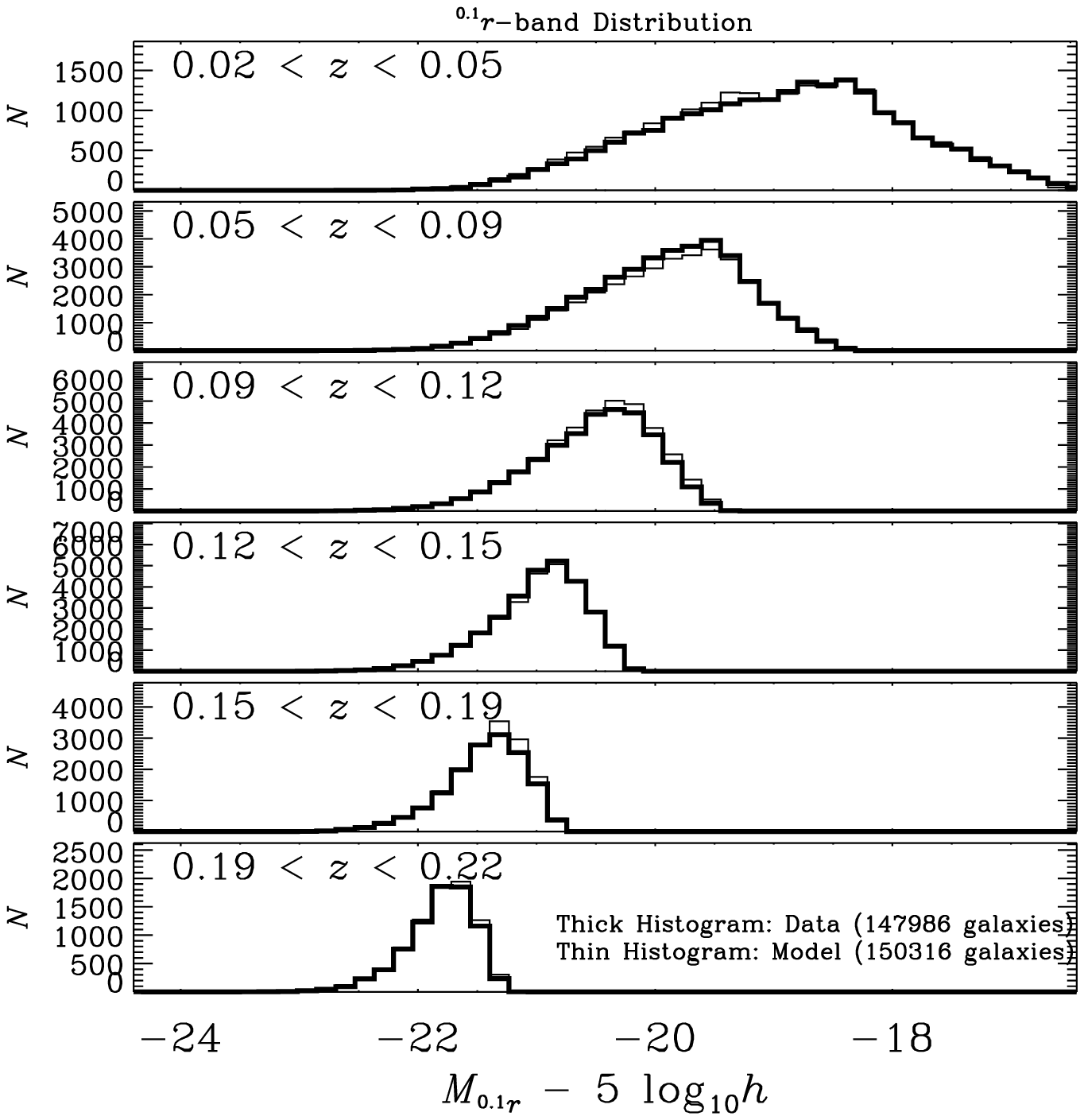}
\caption{\label{rqa} Same as Figure
\ref{uqa}, for the \band{0.1}{r} band.}
\end{figure}

\begin{figure}
\figurenum{\fignum}
\plotone{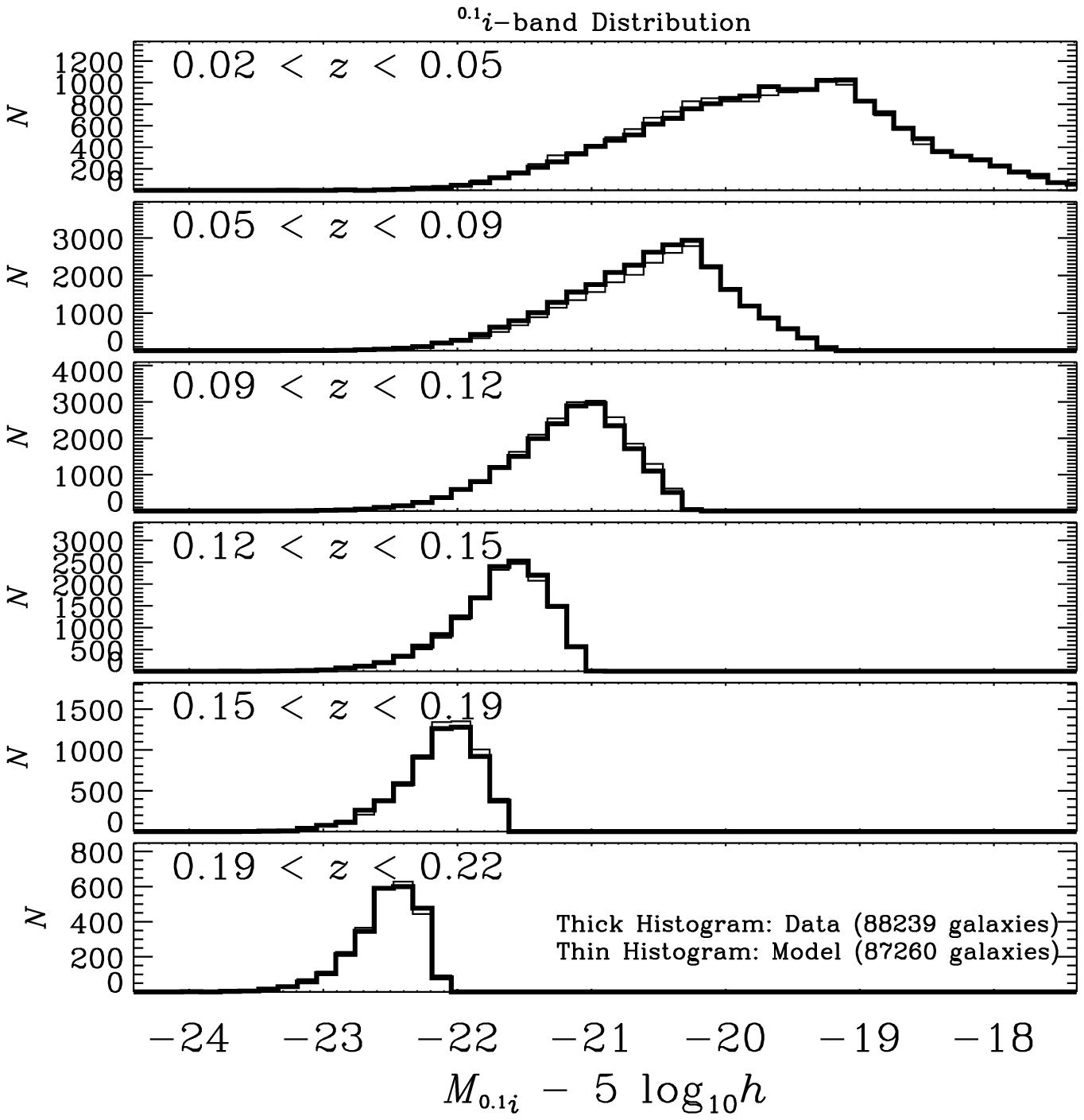}
\caption{\label{iqa} Same as Figure
\ref{uqa}, for the \band{0.1}{i} band.}
\end{figure}

\begin{figure}
\figurenum{\fignum}
\plotone{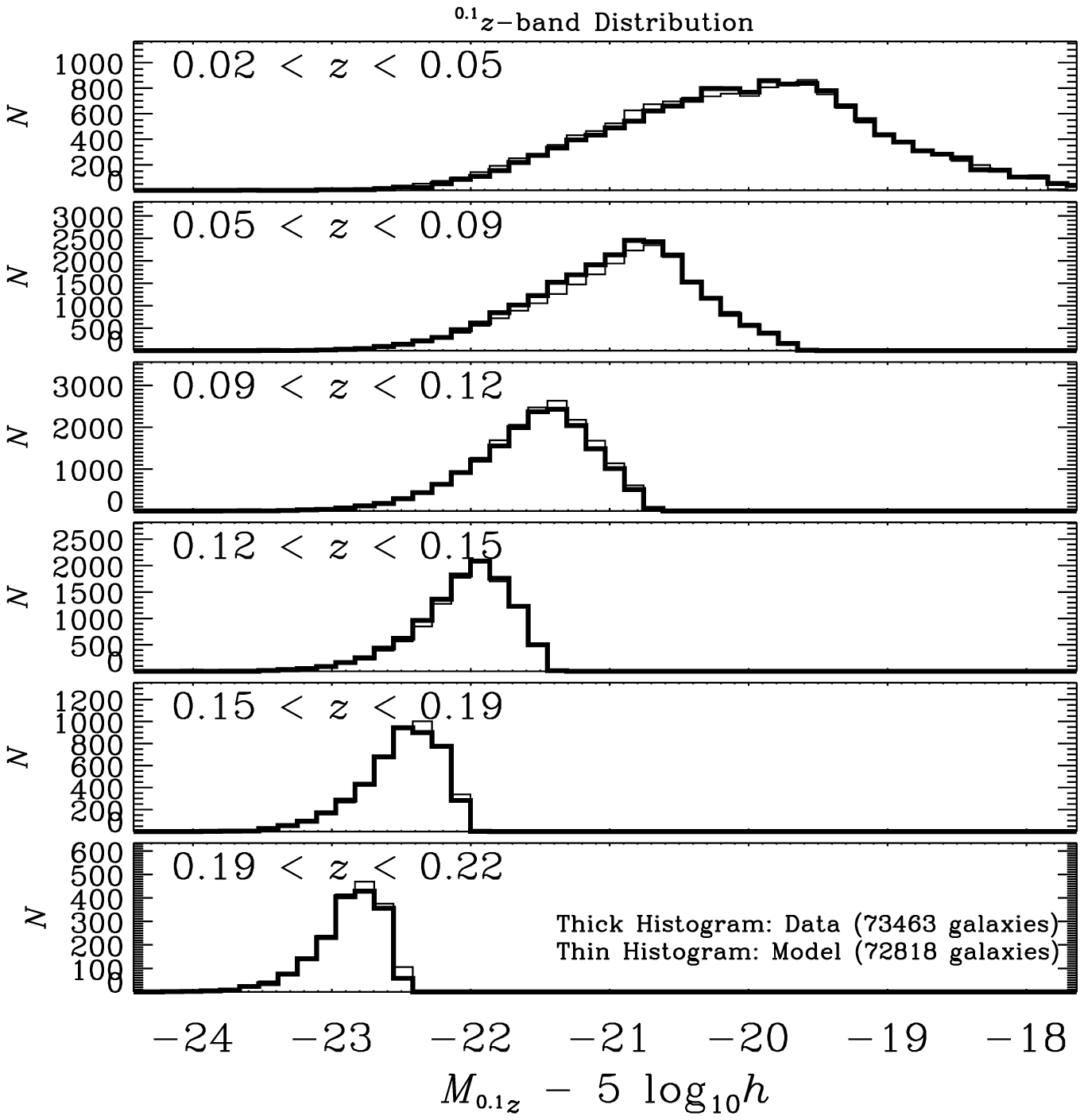}
\caption{\label{zqa} Same as Figure
\ref{uqa}, for the \band{0.1}{z} band.}
\end{figure}

\begin{figure}
\figurenum{\fignum}
\plotone{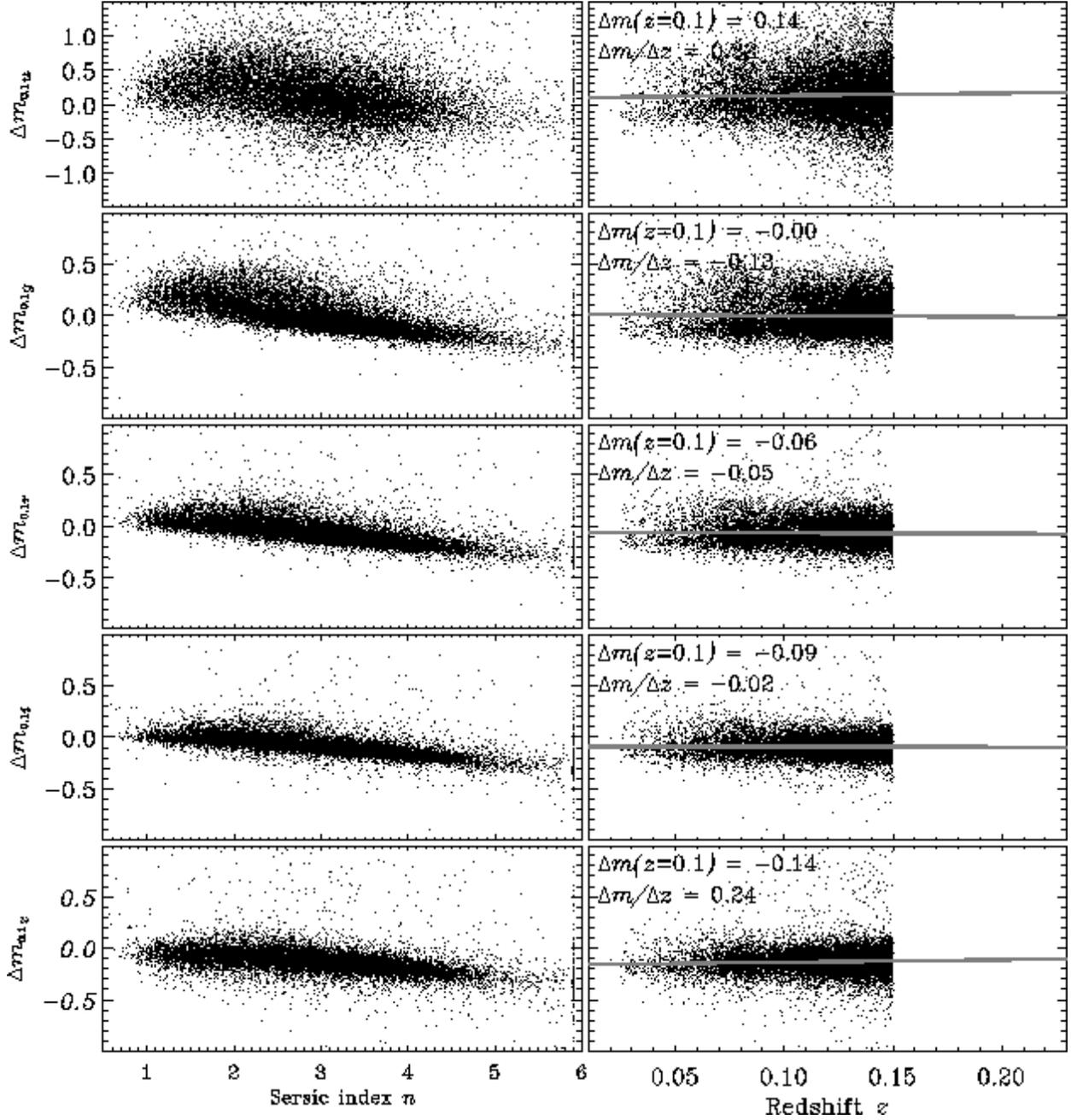}
\caption{\label{check_sersic} Differences $\Delta m \equiv m_S - m_P$
between Sersic and Petrosian magnitudes for each band for a
volume-limited sample with $-23 <M_{\band{0.1}{r}} < -21$ and
$0.02<z<0.15$. The left column shows the differences as a function of
Sersic index $n$. As expected, for galaxies at high Sersic index
(close to the de Vaucouleurs value $n=4$) the Petrosian magnitudes are
an underestimate relative to Sersic magnitudes. For $r$, $i$, and $z$,
the differences between the two remain small at low Sersic index (near
the exponential value $n=1$). The right column shows the differences
as a function of redshift $z$. A linear regression is shown as a grey
line, along with the parameters associated with the best-fit
regression. The slopes are generally insignificant compared to our
uncertainties in the evolution parameter $Q$.}
\end{figure}

Table \ref{kuniv} makes clear that there are large differences between
our present results and those of \citet{blanton01a}. We should
emphasize here that if we apply the same methods used in that paper to
our current sample, the results are within 1$\sigma$ of the results in
that paper. This change in the luminosity density estimates are not
due to statistical fluctuations, nor to a change in the photometric
calibration, but are instead due to improvements in our model for the
luminosity function and the observations.  In particular, we are using
a better estimate of the $K$-corrections and we are including
evolution in our model for the luminosity function. The largest
changes are the luminosity densities in the \band{0.0}{u} and
\band{0.0}{g} bands (which are 1.2 and 0.75 magnitudes less luminous
respectively), where the changes in the $K$-corrections were largest
{\it and} where evolution is most important. Ignoring evolution in the
luminosity function model causes a large bias in the estimate of the
luminosity density, because it causes the expected number of objects
at high redshift to be inaccurate.  If galaxies in fact are brighter
in the past, a non-evolving model tends to yield lower number counts
at high redshift, at a given normalization. Since the normalization
procedure of \citet{davis82a} weights according to volume, and thus
accords higher weight at higher redshift, in this case a non-evolving
model would result in an overestimate of galaxies at low redshift. Due
to bad luck, the systematics comparison of Figure 8 in that paper,
which compared the normalizations of the luminosity function at high
and low redshift, happened not to reveal this effect, presumably due
to the large supercluster at $z \sim 0.08$ in those data (and still
distinctly visible in Figure \ref{zhist} in this, much larger,
dataset!). Figures \ref{zhist}--\ref{zqa} in the present paper show
decisively that our current model explains the redshift counts very
well.

So how does this affect our comparisons to other surveys? For the
LCRS, whose method of fitting the luminosity function and its
normalization was identical to that of \citet{blanton01a}, the
original comparison remains the fair one. That is, even though our
estimate of the luminosity density is now only $0.2$ magnitudes
brighter than that of \citet{lin96a}, this is only an accident,
resulting from a combination of two effects in the LCRS: using bright
isophotal magnitudes, which lowers the luminosity density estimate,
and ignoring (as \citealt{blanton01a} also did) evolution, which
raises the luminosity density estimate. For 2MASS, the change in our
result makes the SDSS more compatible with the results of
\citet{cole01a}. However, it is more difficult to directly compare
these surveys, since the SDSS bands and 2MASS bands do not overlap.

For the 2dFGRS, \citet{norberg02a} report a luminosity density of
$j_{b_j} = -15.35$ absolute magnitudes at $z=0$ (integrating the
Schechter function for the $\Omega_0=0.3$, $\Omega_\Lambda=0.7$
cosmology in the first line of their Table 2 over all luminosities).
This result is based on extrapolating to $z=0$ the luminosities of
galaxies whose typical redshifts are $z=0.05$--$0.2$, using
assumptions about the luminosity evolution. Figure 8 of
\citet{norberg02a}, which shows the mean $K$-correction and evolution
correction used in their analysis, shows that their evolution
correction corresponds closely to $Q=1$. Since we find a somewhat
different value of $Q\sim 2$ at these wavelengths, and both surveys
have similar median redshifts, the fair comparison of the luminosity
densities involves evaluating the luminosity density at around
$z=0.1$.  For this reason, we evolution-correct their results back to
$z=0.1$ by applying $\Delta m = -0.1 Q = -0.1$. Thus, for 2dFGRS
$j_{b_j}(z=0.1) = -15.45 \pm 0.1$, within 1$\sigma$ of our result in
Table \ref{kuniv}.  Note that if we instead compare our $z=0$ value of
the $b_j$ luminosity density to theirs, the discrepancy is about 0.2
magnitudes. However, in either comparison the differences between the
SDSS and 2dFGRS luminosity densities are rather small.

We find similarly small differences between our results and the galaxy
counts data from the Millenium Galaxy Catalog of \citet{liske02a}. Our
Schechter function model for the $\band{0.1}{g}$-band luminosity
function, crudely transformed into $B_{\mathrm{MGC}}$ bandpass by
applying $-0.09$ to $M_\ast$, and using $Q=2$ as for \band{0.1}{g},
nicely predicts their galaxy counts to $B_{\mathrm{MGC}} = 20$ to
within about 5\%.

The original \citet{folkes99a} 2dFGRS result that \citet{blanton01a}
compared to did not evolution-correct their magnitudes either. So why
was there a discrepancy between those two results?  First of all, as
\citet{norberg02a} have pointed out, the linear transformation used by
\citet{blanton01a} to convert SDSS magnitudes to 2dFGRS magnitudes
(which was based on the linear transformation between $b_j$ and $B$
and $V$ published by \citealt{metcalfe95a}) was inappropriate.  Using
the \citet{blair82a} transformation from $B$ and $V$ to $b_j$ ($b_j-B
= -0.28 (B-V)$) instead results in a 10\% reduction in the luminosity
density, not enough to explain the original discrepancy.  However, one
significant difference between \citet{folkes99a} and the mock $b_j$
luminosity function in \citet{blanton01a} was that \citet{folkes99a}
normalized to the number counts instead of using a volume-weighted
method.  Due to this normalization, the \citet{folkes99a} result was
less affected by luminosity evolution, because their normalization was
set by galaxies close to the median redshift of the survey, rather
than those at the distant edge of the survey. This difference seems to
account for the bulk of the discrepancy between \citet{blanton01a} and
\citet{folkes99a}. \citet{norberg02a} reach the same conclusion. As
stated above, comparison of our results to those of
\citet{norberg02a}, who perform an evolution correction to their
magnitudes, now results in a consistent measurement of the luminosity
density.

Nevertheless, \citet{blanton01a} did show that the isophotal
magnitudes meaured by the APM survey did not include a significant
fraction of the flux for a typical galaxy, and further that the
``Gaussian-correction'' to total magnitudes (described in detail by
\citealt{maddox90a}) performed on the galaxy fluxes were insufficient
to replace this difference. If this were the whole story, there should
remain a difference of roughly 30\% between the \citet{norberg02a}
results based on 2dFGRS magnitudes and the results presented here,
based on SDSS Petrosian magnitudes. However, extensive further
calibration of the plate magnitudes has been performed using deep CCD
imaging, as described by \citet{norberg02a}. In effect, the overall
calibration of the galaxy survey is set by a comparison between
observations of a set of galaxies using the APM and using deep CCD
imaging. Since the magnitudes in the deep CCD imaging count flux
outside the limiting isophotes of the APM magnitudes, this overall
calibration translates the isophotal to total magnitudes (on average,
that is --- naturally, for any particular galaxy the correction
depends on the profile shape and size of that galaxy).  For this
reason, the comparisons in \citet{norberg02a} of 2dFGRS magnitudes and
SDSS magnitudes agree on average, for galaxies near the median surface
brightness (although there is a strong correlation of the residuals
with surface brightness). \citet{blanton01a} did not account for these
extra step of calibration and isophotal-to-total correction, and so
overestimated the effect of the different magnitude definitions in the
final luminosity density.

Both \citet{liske02a} and \citet{cole01a} have suggested that the
results of \citet{blanton01a} could have been biased high due to
large-scale structure.  This possibility can be ruled out by simply
using the same methods used by that paper on the full sample presented
here. The luminosity density so calculated is within $1\sigma$ of the
original result, suggesting that the 10\% uncertainties calculated for
that sample were realistic. This agreement occurs despite the fact
that the galaxy counts for $r<17.7$ in the region used by
\citet{blanton01a} are 10--15\% higher than in the survey on
average. As correctly pointed out by \citet{norberg02a}, the
statistical uncertainties in the normalization are much smaller when
using the volume-weighting method of \citet{davis82a}, rather than the
normalizing to galaxy counts, though as those authors state one is
more susceptible to systematic errors in the model for
evolution. Using the luminosity function model in this paper,
normalizing either way yields nearly identical results; for example,
consider the counts predicted by the models listed in Figures
\ref{uqa}--\ref{zqa}. The possible exception is the $u$-band, for
which there is a 10\% difference (still within the $1\sigma$
statistical uncertainties). Again, we emphasize that the differences
between the luminosity function derived here and in \citet{blanton01a}
are a product primarily of the model of the luminosity function we
use.

In short, the discrepancies between SDSS and 2dFGRS appear to be
resolved, for the most part due to a proper treatment of evolution in
the SDSS luminosity function. It remains to be determined whether the
results of SDSS and 2MASS are discrepant, but in any case they are
ameliorated by the reduction of our estimate of the luminosity
density.

\subsection{Luminosity Density as a Function of Wavelength}
 
\begin{figure}
\figurenum{\fignum}
\plotone{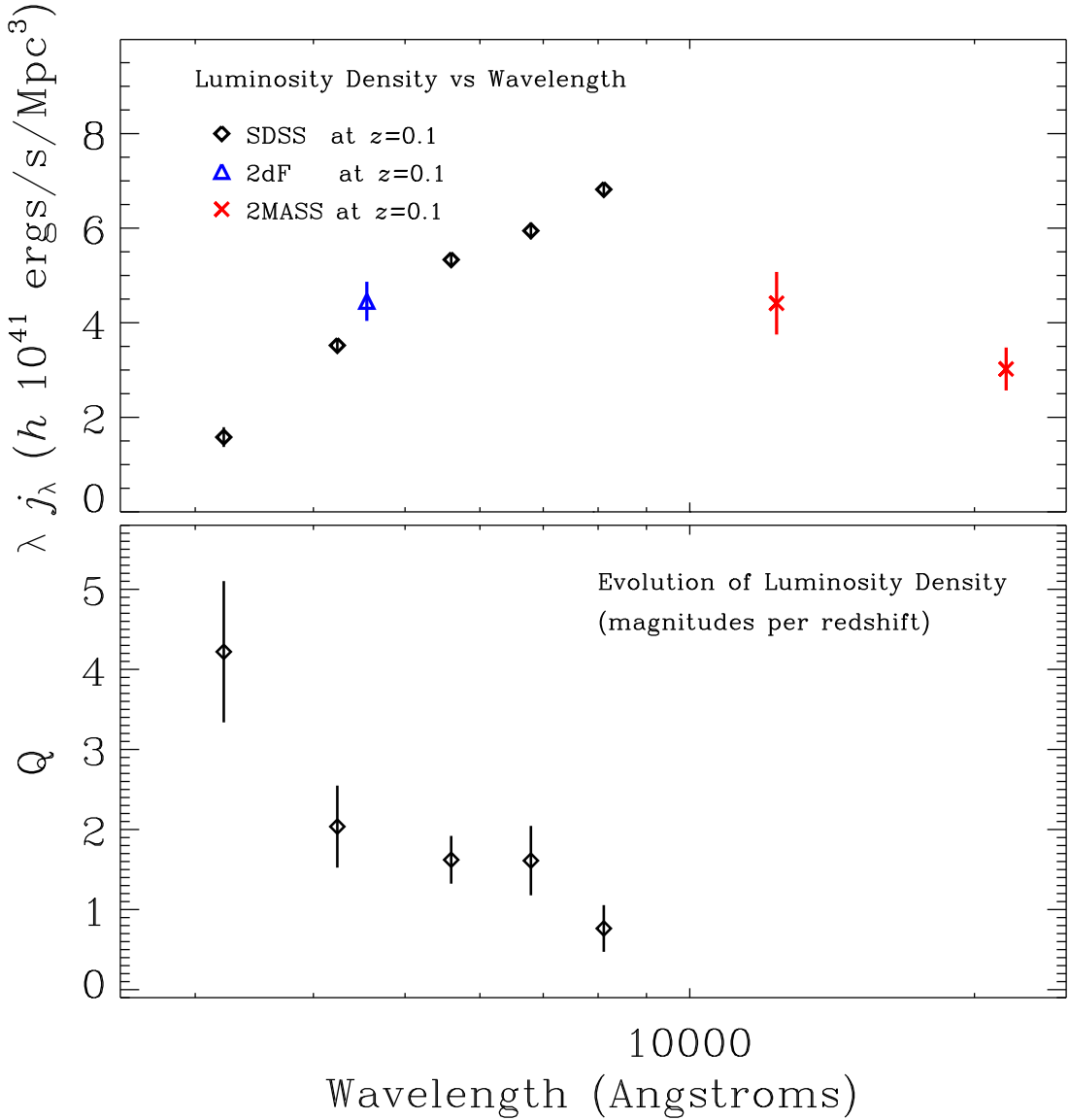}
\caption{\label{mdenq} Luminosity density and its evolution as a
function of wavelength, for the SDSS (diamonds; this paper), the
2dFGRS (triangle; \citealt{norberg02a}), and 2MASS (crosses,
\citealt{cole01a}).  The 2dFGRS has been evolution-corrected to
$z=0.1$ using $Q_{b_j} = 1$ (the effective value used by
\citealt{norberg02a}) and 2MASS has been evolution-corrected to
$z=0.1$ using $Q_J=Q_K=1$. }
%For reference (and only for reference) we
% also show an 8 Gyr old instantaneous burst population from the
%GISSEL96 models (\citealt{bruzual93a}) with a metallicity
%$Z=0.008$. This model is not, of course, a good fit to the data;
%however, it demonstrates that the $J$ and $K$ results of
%\citet{cole01a} are not in obvious disagreement with the SDSS results
%of this paper.}
\end{figure}

Figure \ref{mdenq} plots the luminosity density versus wavelength
using Equation (\ref{fluxeq}) for the SDSS (diamonds). Using the above
results, we add the 2dFGRS point from \citet{norberg02a} (triangle),
and the 2MASS points from \citet{cole01a} (crosses), as described in
this section.

For the 2dFGRS, translating the luminosity density expressed in
magnitudes into physical units is a nontrivial endeavour. The
zeropoint of the 2dFGRS $b_j$ system is set by the equation:
\begin{equation}
\label{bjeq}
b_j - B = -0.28 (B-V).
\end{equation}
Although this slope was determined by \citet{blair82a} based only on a
small set of stars, it can be checked {\it a posteriori} against CCD
observations which are appropriately converted from their natural
system into the Johnson $B$ and $V$ system. According to
\citet{norberg02a}, such observations confirm the slope $-0.28$ to a
surprising degree of accuracy given the estimated errors of
\citet{blair82a}. However, the typical galaxy used to perform this
check is not anywhere near $B-V = 0$, and because we expect the
relationship between any set of bandpasses to be nonlinear (with or
without photographic plate nonlinearities), it is not guaranteed that
$b_j$, as zeropointed using Equation (\ref{bjeq}, is a Vega-relative
magnitude. We perform a test of Equation (\ref{bjeq}) using an
estimate of the spectral response of the plates, from Hewett \& Warren
(private communication), who have determined the response for one
plate by dispersing a spectrum (with well-measured
$f_\lambda(\lambda)$ from spectrophotometry) through 2mm of GG395
filter and onto Kodak IIIaJ emulsion, the definition of the $b_j$
bandpass. Using the nominal $B$ and $V$ responses derived by
\citet{bessell90a} and typical galaxy spectra at $z=0$, we find that
the relationship between $b_j$ and $B$ around a typical galaxy color
of $B-V \sim 0.8$ is close to $b_j - B\approx 0.04 - 0.28 (B-V)$. The
relationship is indeed nonlinear and appears to cross $b_j - B = 0$ at
$B-V = 0$. Again, if this difference existed in the 2dFGRS data, it
would never be noticed because the zeropoint is explicitly set using
CCD imaging of galaxies assuming Equation (\ref{bjeq}) holds at
typical galaxy colors. On the other hand, the curve given to us by
Hewett \& Warren is based only on measurements of a single plate,
whose response may not be typical of the plates in the APM. Given this
uncertainty, we simply assume Hewett \& Warren are correct, and derive
an AB magnitude from 2dFGRS $b_j$ magnitudes:
\begin{equation}
\label{twodftoab}
b_j(\mathrm{AB}) = b_j(\mathrm{2dFGRS}) + 0.04 - 0.07,
\end{equation}
which we claim is uncertain at the 0.05 mag level. Then we can use
Equation (\ref{fluxeq}) to calculate the luminosity density in
physical units.\footnote{This procedure yields a conversion between
magnitudes and physical units substantially different (by about 50\%)
than what one would infer about the conversion from Table 1 of
\citet{folkes99a}.} 

%In passing, we note that applying this procedure to the Schechter
%function quoted in Table 1 of \citet{folkes99a} yields a luminosity
%density in ergs s$^{-1}$ Hz$^{-1}$ Mpc$^{-3}$ 1.5 times higher than
%quoted in the last column of their Table 1. \citet{wright01a} appears
%to have used a value similar to ours in creating his Figure 1,
%implicitly correcting \citet{folkes99a}. This difference is much
%larger than can be explained by using different $b_j$
%passbands. However, since they do not describe how they calculate this
%number, we cannot determine the source of the discrepancy.

In the near infrared, the Two-Micron All Sky Survey (2MASS) results of
\citet{cole00a} for the $\band{0.0}{J}$ and $\band{0.0}{K_s}$ bands
are shown. We use the published responses of $J$ and $K_s$ from the
2MASS website\footnote{{\tt
http://www.ipac.caltech.edu/2mass/releases/second/doc/sec3\_1b1.tbl12.html}}
and the \citet{kurucz91a} theoretical Vega spectrum (normalized at
5000 \AA\ to match the \citet{hayes85a} spectrophotometry of Vega) to
calculate zeropoint shifts from Vega to AB magnitudes of ($0.91$,
$1.85$) for $J$ and $K_s$, respectively. The $J$ band AB zeropoint
corresponds closely with that listed in the 2MASS Explanatory
Supplement Section IV.5.a.  However, the $K_s$ band AB zeropoint
listed there is about 1.77; since the spectrophotometry or model used
to calculate this zeropoint is not specified on that site, we cannot
determine the source of this discrepancy. However, the difference is
within the 1$\sigma$ statistical uncertainties of \citet{cole01a}. To
evolution-correct the 2MASS luminosity densities to $z=0.1$, we assume
$Q=1$, which is consistent with most stellar population synthesis
models. We display this in Figure \ref{mdenq} as crosses.

We should note here that although we have done our best here to place
all the observations on the same physical footing, absolute
calibration of this sort is uncertain. No spectrophotometry we know of
has been performed to verify the models of \citet{kurucz91a} in the
infrared, which we rely on to put the 2MASS results in physical units
in Figure \ref{mdenq}. There are uncertainties of at least 5\% percent
in the spectrophotometric calibration of BD+17 4708, the primary
standard used to calibrate the SDSS. In any of the observations there
is uncertainty and quite possibly variability in the bandpass
responses. So this plot (and any plot like it in the literature)
should not be trusted to better than 5\% (which can be enough to
substantially change one's theoretical interpretation of the
observations in terms of a star-formation history).

%For reference (and only for reference) we also show an 8 Gyr old
%instantaneous burst population from the GISSEL96 models
%(\citealt{bruzual93a}) with a metallicity $Z=0.008$. This model is
%not, of course, a good fit to the data. However, it demonstrates that
%the $\band{0.0}{K}$-band results of \citet{cole01a} are not in obvious
%disagreement with the SDSS results of this paper. Whether the
%remaining discrepancy in the \band{0.0}{J} band is a statistical
%fluctuation, or reflects a large remaining difference in the optical
%and infrared photometry of galaxies, or can be simply explained
%theoretically is uncertain at the moment.  

We leave until a later paper or to other authors a serious attempt to
reconcile these various observations given a star-formation history
and stellar population synthesis models.

% 6.79 5.44 4.76 4.58 4.51 

\section{Conclusions and Future Work}
\label{conclusions}

We have presented the galaxy luminosity density across the optical
range, and have shown that it is consistent with other
determinations. Previous differences found between the SDSS and other
surveys are primarily attributable to the effect of luminosity
evolution on our results. We have accounted for luminosity evolution
in our current fits, and have given our results, which appear loosely
consistent with the predictions of passive evolution in most
bands. Our results are similar to those found by \citet{bernardi02q}
for their sample of early-type galaxies. 

We note that our results are in general agreement with the shape of
the cosmic spectrum determined from the 2dFGRS and SDSS galaxy spectra
by \citet{baldry02a} and Glazebrook et al., in preparation. In the
case of SDSS spectra, Glazebrook et al. use the same data as used here
(Petrosian magnitudes in {\tt sample10}) to apply an overall
normalization to their spectrum, but use the $3''$ diameter fiber
spectra (which typically contain about 20\% of the Petrosian galaxy
flux and cover regions of the galaxy about 0.1 mag redder in $g-r$
than that covered by the Petrosian aperture) to find the luminosity
density as a function of wavelength. The power of their approach is
that the emission and absorption lines contain more detailed
information on the star-formation history than do broad-band colors;
the question remains whether this star-formation history is
representative of the global star-formation history. 

We will continue to improve our estimates of evolution. First, we will
quantify our uncertainties in flux measurements more robustly and
understand how our estimates of magnitude are affected by seeing in
more detail. Doing so will yield more reliable estimates of luminosity
evolution. Second, different galaxy types are expected to evolve
differently; redder galaxies are generally presumed to be older and
thus more slowly evolving. The different evolution of different galaxy
types can constrain theories of the formation of galaxies, and thus it
is of interest to study this differential evolution.  Furthermore, it
is well known that different galaxy types have different luminosity
functions ({\it e.g.}, \citealt{blanton01a}); thus, it is probable
that the shape of the galaxy luminosity function evolves, not merely
its overall luminosity scale.

One caveat to our results is the existence of surface-brightness
selection effects. \citet{blanton01a} demonstrated that if the SDSS
can truly detect most of the objects which exist down to $\mu_{r,50}
\sim 24$, not much luminosity density ($\sim 5\%$) can exist at lower
surface brightnesses, unless the dependence of luminosity on surface
brightness flattens dramatically below these limits or there is a
sharp upturn in the luminosity function at low
luminosities. \citet{cross01a} demonstrated similar results for
2dFGRS. This result is not necessarily inconsistent with the results
of \citet{oneil00a}, who found that the number density of galaxies
does not decline at low surface brightnesses; however, if the
relationship between luminosity and surface brightness measured in the
SDSS exists in their galaxy sample, the contribution to the luminosity
density of galaxies should decline considerably at low surface
brightness. Unfortunately, that paper and its predecessors do not
evaluate the luminosity density contributed by different ranges of
surface brightness in their sample. 

The possibility remains that the SDSS is very incomplete at
$\mu_{r,50} \sim 24$. Such incompleteness would probably not be due to
low signal-to-noise (see the order-of-magnitude calculation in
\citealt{blanton01a}); it would more likely be due to systematic
difficulties in subtracting the sky background. We find no evidence
for changes in the measured extinction-corrected surface-brightness
distribution at Galactic extinctions varying from 0 to nearly 2 in the
$r$-band (as determined by \citealt{schlegel98a}), which indicates
that the surface brightnesses of most galaxies in the survey are not
close to the surface brightness limit of the survey. In fact, there
are a number of ``objects'' detected at low surface brightness
$\mu_{r,50} > 24$, nearly all of which turn out to be scattered light
or other image defects; thus, such low surface brightness features are
readily detectable. We can test these effects more thoroughly by
searching for simulated galaxies inserted into actual data, and such
tests are currently ongoing.

These results (and the accompanying covariance matrices) can be used
to constrain aspects of the star-formation history of the universe,
and in particular the overall stellar density. In addition, they can
be used to develop a selection function for flux-limited galaxy
redshift surveys selected from the SDSS. They also provide the
state-of-the-art luminosity densities with which to calculate
$\Omega_m$ using measured mass-to-light ratios. Finally, they can be
compared to high redshift estimates of luminosity density in order to
evaluate the evolution of galaxies to high redshift.

\acknowledgments

We thank Paul Hewett for providing an estimate of the $b_j$ spectral
response. We had interesting discussions with Ravi Sheth about this
work. MB acknowledges NASA NAG5-11669 for partial support, and is also
grateful for the hospitality of the Department of Physics and
Astronomy at the State University of New York at Stony Brook, who
kindly provided computing facilities on his frequent visits there.

Funding for the creation and distribution of the SDSS Archive has been
provided by the Alfred P. Sloan Foundation, the Participating
Institutions, the National Aeronautics and Space Administration, the
National Science Foundation, the U.S. Department of Energy, the
Japanese Monbukagakusho, and the Max Planck Society. The SDSS Web site
is {\tt http://www.sdss.org/}.

The SDSS is managed by the Astrophysical Research Consortium (ARC) for
the Participating Institutions. The Participating Institutions are The
University of Chicago, Fermilab, the Institute for Advanced Study, the
Japan Participation Group, The Johns Hopkins University, Los Alamos
National Laboratory, the Max-Planck-Institute for Astronomy (MPIA),
the Max-Planck-Institute for Astrophysics (MPA), New Mexico State
University, University of Pittsburgh, Princeton University, the United
States Naval Observatory, and the University of Washington.

\newpage

\clearpage

\begin{deluxetable}{ccccr}
\tablewidth{0pt}
\tablecolumns{5}
%\tablenum{\tabnum}
\tablecaption{\label{limits} Selection limits }
\tablehead{ Band & Flux Limit & Redshift Limits &
Absolute Magnitude Limits & Number of Galaxies}
\tablecomments{Arbitrarily apparently bright objects were included as
long as they passed the galaxy target selection limits of
\citet{strauss02a}. Absolute magnitude limits are those used for the
($\Omega_0=0.3$, $\Omega_\Lambda =0.7$) case. }
\startdata
\band{0.1}{u} & 
   $  m_{u} < 18.36 $ &
   $  0.02 < z <  0.14 $ &
	 $ -21.93 < M_{\band{0.1}{u}} - 5 \log_{10} h < -15.54 $ &
	 $ 22020 $ \cr
\band{0.1}{g} & 
   $  m_{g} < 17.69 $ &
   $  0.02 < z <  0.17 $ &
	 $ -23.38 < M_{\band{0.1}{g}} - 5 \log_{10} h < -16.10 $ &
	 $ 53999 $ \cr
\band{0.1}{r} & 
   $  m_{r} < 17.79 $ &
   $  0.02 < z <  0.22 $ &
	 $ -24.26 < M_{\band{0.1}{r}} - 5 \log_{10} h < -16.11 $ &
	 $ 147986 $ \cr
\band{0.1}{i} & 
   $  m_{i} < 16.91 $ &
   $  0.02 < z <  0.22 $ &
	 $ -23.84 < M_{\band{0.1}{i}} - 5 \log_{10} h < -17.07 $ &
	 $ 88239 $ \cr
\band{0.1}{z} & 
   $  m_{z} < 16.50 $ &
   $  0.02 < z <  0.22 $ &
	 $ -24.08 < M_{\band{0.1}{z}} - 5 \log_{10} h < -17.34 $ &
	 $ 73463 $ \cr
\enddata
\end{deluxetable}

\begin{deluxetable}{cccccc}
\tablewidth{0pt}
\tablecolumns{6}
%\tablenum{\tabnum}
\tablecaption{\label{schtable} Schechter function fits}
\tablehead{ $\Omega_0$ & $\Omega_\Lambda$ & Band & $\phi_\ast$
($10^{-2} h^{3}$ Mpc$^{-3}$) &
$M_\ast - 5 \log_{10} h$ & $\alpha$ }
\tablecomments{The uncertainties are correlated; see Tables
\ref{ucorrtable}--\ref{zcorrtable} for the correlation matrix of the
uncertainties.}
\startdata
0.3 & 0.7  & \band{0.1}{u} & 
   $  3.05 \pm  0.33 $ &
   $ -17.93 \pm  0.03 $ &
   $ -0.92 \pm  0.07 $ \cr
 &  & \band{0.1}{g} & 
   $  2.18 \pm  0.08 $ &
   $ -19.39 \pm  0.02 $ &
   $ -0.89 \pm  0.03 $ \cr
 &  & \band{0.1}{r} & 
   $  1.49 \pm  0.04 $ &
   $ -20.44 \pm  0.01 $ &
   $ -1.05 \pm  0.01 $ \cr
 &  & \band{0.1}{i} & 
   $  1.47 \pm  0.04 $ &
   $ -20.82 \pm  0.02 $ &
   $ -1.00 \pm  0.02 $ \cr
 &  & \band{0.1}{z} & 
   $  1.35 \pm  0.04 $ &
   $ -21.18 \pm  0.02 $ &
   $ -1.08 \pm  0.02 $ \cr
\smallskip
0.3 & 0.0  & \band{0.1}{u} & 
   $  3.26 \pm  0.40 $ &
   $ -17.89 \pm  0.04 $ &
   $ -0.94 \pm  0.09 $ \cr
 &  & \band{0.1}{g} & 
   $  2.42 \pm  0.10 $ &
   $ -19.34 \pm  0.02 $ &
   $ -0.92 \pm  0.04 $ \cr
 &  & \band{0.1}{r} & 
   $  1.69 \pm  0.06 $ &
   $ -20.37 \pm  0.02 $ &
   $ -1.03 \pm  0.03 $ \cr
 &  & \band{0.1}{i} & 
   $  1.62 \pm  0.06 $ &
   $ -20.78 \pm  0.03 $ &
   $ -1.02 \pm  0.04 $ \cr
 &  & \band{0.1}{z} & 
   $  1.47 \pm  0.05 $ &
   $ -21.12 \pm  0.02 $ &
   $ -1.07 \pm  0.03 $ \cr
\smallskip
1.0 & 0.0  & \band{0.1}{u} & 
   $  3.65 \pm  0.40 $ &
   $ -17.83 \pm  0.04 $ &
   $ -0.90 \pm  0.06 $ \cr
 &  & \band{0.1}{g} & 
   $  2.62 \pm  0.10 $ &
   $ -19.30 \pm  0.02 $ &
   $ -0.91 \pm  0.03 $ \cr
 &  & \band{0.1}{r} & 
   $  1.83 \pm  0.06 $ &
   $ -20.33 \pm  0.03 $ &
   $ -1.04 \pm  0.03 $ \cr
 &  & \band{0.1}{i} & 
   $  1.73 \pm  0.06 $ &
   $ -20.74 \pm  0.02 $ &
   $ -1.03 \pm  0.03 $ \cr
 &  & \band{0.1}{z} & 
   $  1.57 \pm  0.05 $ &
   $ -21.11 \pm  0.02 $ &
   $ -1.10 \pm  0.02 $ \cr
\enddata
\end{deluxetable}

\clearpage
\begin{landscape}

\begin{deluxetable}{cccccccccc}
\tablewidth{0pt}
\tablecolumns{10}
%\tablenum{\tabnum}
\tablecaption{\label{lftable} Luminosity density and evolution
parameters}

\tablehead{ $\Omega_0$ & $\Omega_\Lambda$ & Band &
$\lambda_{\mathrm{eff}}$ & $j + 2.5 \log_{10} h$ & $j$ & $j$ &
$Q$ & $P$ & $f_{\mathrm{np}}$ \\ & & & (\AA) & (mags in Mpc$^3$) &
($h$ $10^8 L_\odot$ Mpc$^{-3}$) & ($h$ $10^{37}$ ergs
cm$^{-2}$Mpc$^{-3}$) }

\tablecomments{$f_{\mathrm{np}}$ is the fraction of the luminosity
density contributed by the nonparametric fit; in principle
$f_{\mathrm{np}}$ can be greater than unity.  The uncertainties are
correlated; see Tables \ref{ucorrtable}--\ref{zcorrtable} for the
correlation matrix of the uncertainties. See Table \ref{fulljk} for
the correlation matrix of the luminosity densities and the luminosity
evolution parameters of all the bands.}  
\startdata
0.3 & 0.7  & \band{0.1}{u} & 
         3216 &
   $ -14.10 \pm  0.15 $ &
   $  2.29 \pm  0.32 $ &
   $  5.48 \pm  0.75 $ &
   $  4.22 \pm  0.88 $ &
   $  3.20 \pm  3.31 $ &
   $  0.90 $ \cr
 &  & \band{0.1}{g} & 
         4240 &
   $ -15.18 \pm  0.03 $ &
   $  1.78 \pm  0.05 $ &
   $  8.54 \pm  0.26 $ &
   $  2.04 \pm  0.51 $ &
   $  0.32 \pm  1.70 $ &
   $  0.97 $ \cr
 &  & \band{0.1}{r} & 
         5595 &
   $ -15.90 \pm  0.03 $ &
   $  1.84 \pm  0.04 $ &
   $  9.57 \pm  0.22 $ &
   $  1.62 \pm  0.30 $ &
   $  0.18 \pm  0.57 $ &
   $  1.00 $ \cr
 &  & \band{0.1}{i} & 
         6792 &
   $ -16.24 \pm  0.03 $ &
   $  2.12 \pm  0.05 $ &
   $  8.82 \pm  0.21 $ &
   $  1.61 \pm  0.43 $ &
   $  0.58 \pm  1.06 $ &
   $  0.99 $ \cr
 &  & \band{0.1}{z} & 
         8111 &
   $ -16.56 \pm  0.02 $ &
   $  2.69 \pm  0.05 $ &
   $  8.33 \pm  0.15 $ &
   $  0.76 \pm  0.29 $ &
   $  2.28 \pm  0.79 $ &
   $  1.01 $ \cr
\smallskip
0.3 & 0.0  & \band{0.1}{u} & 
         3216 &
   $ -14.14 \pm  0.19 $ &
   $  2.39 \pm  0.42 $ &
   $  5.70 \pm  1.00 $ &
   $  3.67 \pm  0.89 $ &
   $  4.02 \pm  3.18 $ &
   $  0.90 $ \cr
 &  & \band{0.1}{g} & 
         4240 &
   $ -15.26 \pm  0.07 $ &
   $  1.92 \pm  0.12 $ &
   $  9.17 \pm  0.55 $ &
   $  1.22 \pm  0.73 $ &
   $  1.82 \pm  2.10 $ &
   $  0.97 $ \cr
 &  & \band{0.1}{r} & 
         5595 &
   $ -15.95 \pm  0.03 $ &
   $  1.93 \pm  0.06 $ &
   $ 10.03 \pm  0.30 $ &
   $  1.11 \pm  0.48 $ &
   $  0.99 \pm  0.96 $ &
   $  1.00 $ \cr
 &  & \band{0.1}{i} & 
         6792 &
   $ -16.31 \pm  0.04 $ &
   $  2.27 \pm  0.09 $ &
   $  9.47 \pm  0.36 $ &
   $  0.94 \pm  0.46 $ &
   $  1.71 \pm  1.14 $ &
   $  0.98 $ \cr
 &  & \band{0.1}{z} & 
         8111 &
   $ -16.59 \pm  0.04 $ &
   $  2.75 \pm  0.09 $ &
   $  8.52 \pm  0.29 $ &
   $  0.48 \pm  0.48 $ &
   $  2.54 \pm  1.23 $ &
   $  1.01 $ \cr
\smallskip
1.0 & 0.0  & \band{0.1}{u} & 
         3216 &
   $ -14.18 \pm  0.14 $ &
   $  2.48 \pm  0.33 $ &
   $  5.94 \pm  0.78 $ &
   $  3.33 \pm  0.66 $ &
   $  4.89 \pm  2.67 $ &
   $  0.90 $ \cr
 &  & \band{0.1}{g} & 
         4240 &
   $ -15.29 \pm  0.05 $ &
   $  1.99 \pm  0.09 $ &
   $  9.50 \pm  0.42 $ &
   $  0.95 \pm  0.49 $ &
   $  2.52 \pm  1.61 $ &
   $  0.97 $ \cr
 &  & \band{0.1}{r} & 
         5595 &
   $ -16.01 \pm  0.04 $ &
   $  2.03 \pm  0.07 $ &
   $ 10.54 \pm  0.34 $ &
   $  0.63 \pm  0.55 $ &
   $  2.16 \pm  1.07 $ &
   $  1.00 $ \cr
 &  & \band{0.1}{i} & 
         6792 &
   $ -16.36 \pm  0.03 $ &
   $  2.36 \pm  0.08 $ &
   $  9.86 \pm  0.32 $ &
   $  0.60 \pm  0.47 $ &
   $  2.63 \pm  1.21 $ &
   $  0.98 $ \cr
 &  & \band{0.1}{z} & 
         8111 &
   $ -16.67 \pm  0.03 $ &
   $  2.98 \pm  0.08 $ &
   $  9.24 \pm  0.24 $ &
   $ -0.29 \pm  0.29 $ &
   $  4.44 \pm  0.79 $ &
   $  1.01 $ \cr
\enddata
\end{deluxetable}

\end{landscape}

\begin{deluxetable}{rrrrrrr}
\tablewidth{0pt}
\tablecolumns{7}
%\tablenum{\tabnum}
\tablecaption{\label{ucorrtable} \band{0.1}{u} band uncertainty correlation
matrix}
\tablehead{ & \graybox{$\delta j_M$} & $\delta Q$ & 
\graybox{$\delta P$} & $\delta \phi_\ast$ &
\graybox{$\delta M_\ast$} & $\delta \alpha$ }
\tablecomments{Correlation matrix calculated from 30 jackknife
resamplings of the data, as described in the text. 
This correlation matrix determined
from the $\Omega_0=0.3$, $\Omega_\Lambda=0.7$ cosmology.}
\startdata
$ \delta j_M $  & $ \graybox{ 1.000} $  & $  0.949 $  & $ \graybox{-0.934} $  & $ -0.938 $  & $ \graybox{-0.091} $  & $  0.746 $ \cr
$ \delta Q $  & $ \graybox{ 0.949} $  & $  1.000 $  & $ \graybox{-0.955} $  & $ -0.834 $  & $ \graybox{ 0.042} $  & $  0.823 $ \cr
$ \delta P $  & $ \graybox{-0.934} $  & $ -0.955 $  & $ \graybox{ 1.000} $  & $  0.837 $  & $ \graybox{-0.013} $  & $ -0.802 $ \cr
$ \delta \phi_\ast $  & $ \graybox{-0.938} $  & $ -0.834 $  & $ \graybox{ 0.837} $  & $  1.000 $  & $ \graybox{ 0.425} $  & $ -0.484 $ \cr
$ \delta M_\ast $  & $ \graybox{-0.091} $  & $  0.042 $  & $ \graybox{-0.013} $  & $  0.425 $  & $ \graybox{ 1.000} $  & $  0.560 $ \cr
$ \delta \alpha $  & $ \graybox{ 0.746} $  & $  0.823 $  & $ \graybox{-0.802} $  & $ -0.484 $  & $ \graybox{ 0.560} $  & $  1.000 $ \cr
\enddata
\end{deluxetable}

\begin{deluxetable}{rrrrrrr}
\tablewidth{0pt}
\tablecolumns{7}
%\tablenum{\tabnum}
\tablecaption{\label{gcorrtable} \band{0.1}{g} band uncertainty correlation
matrix}
\tablecomments{Correlation matrix calculated from 30 jackknife
resamplings of the data, as described in the text. 
This correlation matrix determined
from the $\Omega_0=0.3$, $\Omega_\Lambda=0.7$ cosmology.}
\tablehead{ & \graybox{$\delta j_M$} & $\delta Q$ & 
\graybox{$\delta P$} & $\delta \phi_\ast$ &
\graybox{$\delta M_\ast$} & $\delta \alpha$ }
\startdata
$ \delta j_M $  & $ \graybox{ 1.000} $  & $  0.930 $  & $ \graybox{-0.885} $  & $ -0.701 $  & $ \graybox{-0.093} $  & $  0.493 $ \cr
$ \delta Q $  & $ \graybox{ 0.930} $  & $  1.000 $  & $ \graybox{-0.949} $  & $ -0.494 $  & $ \graybox{ 0.130} $  & $  0.584 $ \cr
$ \delta P $  & $ \graybox{-0.885} $  & $ -0.949 $  & $ \graybox{ 1.000} $  & $  0.447 $  & $ \graybox{-0.144} $  & $ -0.656 $ \cr
$ \delta \phi_\ast $  & $ \graybox{-0.701} $  & $ -0.494 $  & $ \graybox{ 0.447} $  & $  1.000 $  & $ \graybox{ 0.766} $  & $  0.219 $ \cr
$ \delta M_\ast $  & $ \graybox{-0.093} $  & $  0.130 $  & $ \graybox{-0.144} $  & $  0.766 $  & $ \graybox{ 1.000} $  & $  0.760 $ \cr
$ \delta \alpha $  & $ \graybox{ 0.493} $  & $  0.584 $  & $ \graybox{-0.656} $  & $  0.219 $  & $ \graybox{ 0.760} $  & $  1.000 $ \cr
\enddata
\end{deluxetable}

\begin{deluxetable}{rrrrrrr}
\tablewidth{0pt}
\tablecolumns{7}
%\tablenum{\tabnum}
\tablecaption{\label{rcorrtable} \band{0.1}{r} band uncertainty correlation
matrix}
\tablecomments{Correlation matrix calculated from 30 jackknife
resamplings of the data, as described in the text. 
This correlation matrix determined
from the $\Omega_0=0.3$, $\Omega_\Lambda=0.7$ cosmology.}
\tablehead{ & \graybox{$\delta j_M$} & $\delta Q$ & 
\graybox{$\delta P$} & $\delta \phi_\ast$ &
\graybox{$\delta M_\ast$} & $\delta \alpha$ }
\startdata
$ \delta j_M $  & $ \graybox{ 1.000} $  & $  0.724 $  & $ \graybox{-0.295} $  & $ -0.644 $  & $ \graybox{ 0.325} $  & $  0.053 $ \cr
$ \delta Q $  & $ \graybox{ 0.724} $  & $  1.000 $  & $ \graybox{-0.849} $  & $ -0.285 $  & $ \graybox{ 0.460} $  & $  0.222 $ \cr
$ \delta P $  & $ \graybox{-0.295} $  & $ -0.849 $  & $ \graybox{ 1.000} $  & $ -0.048 $  & $ \graybox{-0.364} $  & $ -0.296 $ \cr
$ \delta \phi_\ast $  & $ \graybox{-0.644} $  & $ -0.285 $  & $ \graybox{-0.048} $  & $  1.000 $  & $ \graybox{ 0.498} $  & $  0.654 $ \cr
$ \delta M_\ast $  & $ \graybox{ 0.325} $  & $  0.460 $  & $ \graybox{-0.364} $  & $  0.498 $  & $ \graybox{ 1.000} $  & $  0.866 $ \cr
$ \delta \alpha $  & $ \graybox{ 0.053} $  & $  0.222 $  & $ \graybox{-0.296} $  & $  0.654 $  & $ \graybox{ 0.866} $  & $  1.000 $ \cr
\enddata
\end{deluxetable}

\begin{deluxetable}{rrrrrrr}
\tablewidth{0pt}
\tablecolumns{7}
%\tablenum{\tabnum}
\tablecaption{\label{icorrtable} \band{0.1}{i} band uncertainty correlation
matrix}
\tablecomments{Correlation matrix calculated from 30 jackknife
resamplings of the data, as described in the text. 
This correlation matrix determined
from the $\Omega_0=0.3$, $\Omega_\Lambda=0.7$ cosmology.}
\tablehead{ & \graybox{$\delta j_M$} & $\delta Q$ & 
\graybox{$\delta P$} & $\delta \phi_\ast$ &
\graybox{$\delta M_\ast$} & $\delta \alpha$ }
\startdata
$ \delta j_M $  & $ \graybox{ 1.000} $  & $  0.889 $  & $ \graybox{-0.763} $  & $ -0.195 $  & $ \graybox{ 0.495} $  & $  0.568 $ \cr
$ \delta Q $  & $ \graybox{ 0.889} $  & $  1.000 $  & $ \graybox{-0.950} $  & $  0.021 $  & $ \graybox{ 0.599} $  & $  0.607 $ \cr
$ \delta P $  & $ \graybox{-0.763} $  & $ -0.950 $  & $ \graybox{ 1.000} $  & $ -0.131 $  & $ \graybox{-0.574} $  & $ -0.654 $ \cr
$ \delta \phi_\ast $  & $ \graybox{-0.195} $  & $  0.021 $  & $ \graybox{-0.131} $  & $  1.000 $  & $ \graybox{ 0.735} $  & $  0.616 $ \cr
$ \delta M_\ast $  & $ \graybox{ 0.495} $  & $  0.599 $  & $ \graybox{-0.574} $  & $  0.735 $  & $ \graybox{ 1.000} $  & $  0.905 $ \cr
$ \delta \alpha $  & $ \graybox{ 0.568} $  & $  0.607 $  & $ \graybox{-0.654} $  & $  0.616 $  & $ \graybox{ 0.905} $  & $  1.000 $ \cr
\enddata
\end{deluxetable}

\begin{deluxetable}{rrrrrrr}
\tablewidth{0pt}
\tablecolumns{7}
%\tablenum{\tabnum}
\tablecaption{\label{zcorrtable} \band{0.1}{z} band uncertainty correlation
matrix}
\tablecomments{Correlation matrix calculated from 30 jackknife
resamplings of the data, as described in the text. 
This correlation matrix determined
from the $\Omega_0=0.3$, $\Omega_\Lambda=0.7$ cosmology.}
\tablehead{ & \graybox{$\delta j_M$} & $\delta Q$ & 
\graybox{$\delta P$} & $\delta \phi_\ast$ &
\graybox{$\delta M_\ast$} & $\delta \alpha$ }
\startdata
$ \delta j_M $  & $ \graybox{ 1.000} $  & $  0.328 $  & $ \graybox{-0.110} $  & $ -0.312 $  & $ \graybox{ 0.086} $  & $  0.249 $ \cr
$ \delta Q $  & $ \graybox{ 0.328} $  & $  1.000 $  & $ \graybox{-0.908} $  & $ -0.265 $  & $ \graybox{ 0.085} $  & $  0.028 $ \cr
$ \delta P $  & $ \graybox{-0.110} $  & $ -0.908 $  & $ \graybox{ 1.000} $  & $  0.171 $  & $ \graybox{-0.008} $  & $ -0.053 $ \cr
$ \delta \phi_\ast $  & $ \graybox{-0.312} $  & $ -0.265 $  & $ \graybox{ 0.171} $  & $  1.000 $  & $ \graybox{ 0.860} $  & $  0.760 $ \cr
$ \delta M_\ast $  & $ \graybox{ 0.086} $  & $  0.085 $  & $ \graybox{-0.008} $  & $  0.860 $  & $ \graybox{ 1.000} $  & $  0.885 $ \cr
$ \delta \alpha $  & $ \graybox{ 0.249} $  & $  0.028 $  & $ \graybox{-0.053} $  & $  0.760 $  & $ \graybox{ 0.885} $  & $  1.000 $ \cr
\enddata
\end{deluxetable}

\clearpage
\begin{landscape}

\begin{deluxetable}{rrrrrrrrrrrr}
\tablewidth{0pt}
\tablecolumns{12}
%\tablenum{\tabnum}
\tablecaption{\label{fulljk} Uncertainty correlation matrix between
luminosity density and evolution fits to all SDSS bands}
\tablecomments{Correlation matrix calculated from 30 jackknife
resamplings of the data, as described in the text.  This correlation
matrix determined from the ($\Omega_0=0.3$, $\Omega_\Lambda=0.7$)
cosmology. We keep enough significant figures that no element of the
inverse of this correlation matrix differs by more than a couple of
percent from that calculated using the machine precision result,
purely to avoid confusion between results we might obtain from a
machine precision version of this matrix and those others might obtain
from this table.  }
\tablehead{ & $\sigma$ &
 \graybox{$\delta j_{\band{0.1}{u}}$} &
$ \delta Q_{\band{0.1}{u}} $ &
\graybox{$ \delta j_{\band{0.1}{g}} $} &
$ \delta Q_{\band{0.1}{g}} $ &
\graybox{$ \delta j_{\band{0.1}{r}} $} &
$ \delta Q_{\band{0.1}{r}} $ &
\graybox{$ \delta j_{\band{0.1}{i}} $} &
$ \delta Q_{\band{0.1}{i}} $ &
\graybox{$ \delta j_{\band{0.1}{z}} $} &
$ \delta Q_{\band{0.1}{z}} $}
\startdata
$ \delta j_{\band{0.1}{u}} $  & $  0.14364 $  & $ \graybox{ 1.00000} $  & $  0.94949 $  & $ \graybox{ 0.57194} $  & $  0.54212 $  & $ \graybox{ 0.44706} $  & $  0.48121 $  & $ \graybox{ 0.33898} $  & $  0.34904 $  & $ \graybox{ 0.11632} $  & $  0.40158 $ \cr
$ \delta Q_{\band{0.1}{u}} $  & $  0.88265 $  & $ \graybox{ 0.94949} $  & $  1.00000 $  & $ \graybox{ 0.58686} $  & $  0.59719 $  & $ \graybox{ 0.43549} $  & $  0.41320 $  & $ \graybox{ 0.26201} $  & $  0.28808 $  & $ \graybox{ 0.10207} $  & $  0.35697 $ \cr
$ \delta j_{\band{0.1}{g}} $  & $  0.03854 $  & $ \graybox{ 0.57194} $  & $  0.58686 $  & $ \graybox{ 1.00000} $  & $  0.92950 $  & $ \graybox{ 0.57137} $  & $  0.51576 $  & $ \graybox{ 0.63139} $  & $  0.60449 $  & $ \graybox{ 0.43418} $  & $  0.71319 $ \cr
$ \delta Q_{\band{0.1}{g}} $  & $  0.51236 $  & $ \graybox{ 0.54212} $  & $  0.59719 $  & $ \graybox{ 0.92950} $  & $  1.00000 $  & $ \graybox{ 0.63424} $  & $  0.54834 $  & $ \graybox{ 0.62528} $  & $  0.64712 $  & $ \graybox{ 0.30430} $  & $  0.71878 $ \cr
$ \delta j_{\band{0.1}{r}} $  & $  0.02843 $  & $ \graybox{ 0.44706} $  & $  0.43549 $  & $ \graybox{ 0.57137} $  & $  0.63424 $  & $ \graybox{ 1.00000} $  & $  0.72439 $  & $ \graybox{ 0.83202} $  & $  0.67494 $  & $ \graybox{ 0.56304} $  & $  0.68807 $ \cr
$ \delta Q_{\band{0.1}{r}} $  & $  0.29843 $  & $ \graybox{ 0.48121} $  & $  0.41320 $  & $ \graybox{ 0.51576} $  & $  0.54834 $  & $ \graybox{ 0.72439} $  & $  1.00000 $  & $ \graybox{ 0.80281} $  & $  0.86152 $  & $ \graybox{ 0.23505} $  & $  0.82870 $ \cr
$ \delta j_{\band{0.1}{i}} $  & $  0.03295 $  & $ \graybox{ 0.33898} $  & $  0.26201 $  & $ \graybox{ 0.63139} $  & $  0.62528 $  & $ \graybox{ 0.83202} $  & $  0.80281 $  & $ \graybox{ 1.00000} $  & $  0.88864 $  & $ \graybox{ 0.59246} $  & $  0.86096 $ \cr
$ \delta Q_{\band{0.1}{i}} $  & $  0.43340 $  & $ \graybox{ 0.34904} $  & $  0.28808 $  & $ \graybox{ 0.60449} $  & $  0.64712 $  & $ \graybox{ 0.67494} $  & $  0.86152 $  & $ \graybox{ 0.88864} $  & $  1.00000 $  & $ \graybox{ 0.21461} $  & $  0.93603 $ \cr
$ \delta j_{\band{0.1}{z}} $  & $  0.02110 $  & $ \graybox{ 0.11632} $  & $  0.10207 $  & $ \graybox{ 0.43418} $  & $  0.30430 $  & $ \graybox{ 0.56304} $  & $  0.23505 $  & $ \graybox{ 0.59246} $  & $  0.21461 $  & $ \graybox{ 1.00000} $  & $  0.32770 $ \cr
$ \delta Q_{\band{0.1}{z}} $  & $  0.29156 $  & $ \graybox{ 0.40158} $  & $  0.35697 $  & $ \graybox{ 0.71319} $  & $  0.71878 $  & $ \graybox{ 0.68807} $  & $  0.82870 $  & $ \graybox{ 0.86096} $  & $  0.93603 $  & $ \graybox{ 0.32770} $  & $  1.00000 $ \cr
\enddata
\end{deluxetable}

\begin{deluxetable}{cccccccc}
\tablewidth{0pt}
\tablecolumns{8}
%\tablenum{\tabnum}
\tablecaption{\label{kuniv} Luminosity density $K$-corrected to
various bandpasses}
\tablecomments{We have taken the luminosity densities in Table
\ref{lftable} for the ($\Omega_0=0.3$, $\Omega_\Lambda$) cosmology and
applied the methods of \citet{blanton02b} in order to evaluate the
luminosity density in a number of other bandpasses. We also show
2dFGRS results from \citet{norberg02a} and old
SDSS results from \citet{blanton01a}. We have inferred the $z=0.1$
value of the luminosity density from \citet{norberg02a} based on the
evolutionary corrections in their Figure 8. $BVRI$ magnitudes assume the
\citet{bessell90a} response curves. We set $b_j = B -0.28 (B-V)$ by
definition (except in the value from \citealt{blanton01a}, where we
copy the number directly from their table). $\Delta m_{AB}$ is the offset
to apply to translate the listed magnitude into an AB magnitude (using
the \citealt{hayes85a} Vega spectrum). } 

\tablehead{ & & & & & $j + 2.5 \log_{10} h$ & & \cr
Band & $\lambda_{\mathrm{eff}}$ (\AA) & $\Delta m_{AB}$ & 
This paper & This paper & SDSS Comm. Data & 2dFGRS & 2dFGRS \cr
& & & $z=0$ & $z=0.1$ & $z=0$ & $z=0$ & $z=0.1$ }
\startdata
\band{0.0}{u} & $     3538 $ & $   0.00 $ & $ -13.99 $ & $ -14.35 $  & $ -15.21 $ & --- & --- \cr
\band{0.0}{g} & $     4664 $ & $   0.00 $ & $ -15.27 $ & $ -15.46 $  & $ -16.05 $ & --- & --- \cr
\band{0.0}{r} & $     6154 $ & $   0.00 $ & $ -15.90 $ & $ -16.06 $  & $ -16.41 $ & --- & --- \cr
\band{0.0}{i} & $     7471 $ & $   0.00 $ & $ -16.29 $ & $ -16.41 $  & $ -16.74 $ & --- & --- \cr
\band{0.0}{z} & $     8922 $ & $   0.00 $ & $ -16.74 $ & $ -16.79 $  & $ -17.02 $ & --- & --- \cr
\band{0.0}{B} & $     3974 $ & $  -0.12 $ & $ -14.98 $ & $ -15.17 $  & --- & --- & --- \cr
\band{0.0}{V} & $     4980 $ & $   0.02 $ & $ -15.69 $ & $ -15.85 $  & --- & --- & --- \cr
\band{0.0}{R} & $     5906 $ & $   0.21 $ & $ -16.23 $ & $ -16.39 $  & --- & --- & --- \cr
\band{0.0}{I} & $     7291 $ & $   0.45 $ & $ -16.90 $ & $ -16.99 $  & --- & --- & --- \cr
\band{0.0}{b_j} & $     4141 $ & $  -0.08 $ & $ -15.18 $ & $ -15.36 $  & $ -15.97 $ & $ -15.35 $ & $ -15.45$  \cr
\enddata
\end{deluxetable}

\end{landscape}

\clearpage
\clearpage

\setcounter{thefigs}{0}

\clearpage
\stepcounter{thefigs}

\clearpage
\stepcounter{thefigs}

\clearpage
\stepcounter{thefigs}

\clearpage
\stepcounter{thefigs}

\clearpage
\stepcounter{thefigs}

\clearpage
\stepcounter{thefigs}

\clearpage
\stepcounter{thefigs}

\clearpage
\stepcounter{thefigs}

\clearpage
\stepcounter{thefigs}

\clearpage
\stepcounter{thefigs}

\clearpage
\stepcounter{thefigs}

\clearpage
\stepcounter{thefigs}

\clearpage
\stepcounter{thefigs}

\clearpage
\stepcounter{thefigs}


\begin{thebibliography}{43}
\expandafter\ifx\csname natexlab\endcsname\relax\def\natexlab#1{#1}\fi

\bibitem[{{Baldry} {et~al.}(2002){Baldry}, {Glazebrook}, {Baugh},
  {Bland-Hawthorn}, {Bridges}, {Cannon}, {Cole}, {Colless}, {Collins}, {Couch},
  {Dalton}, {De Propris}, {Driver}, {Efstathiou}, {Ellis}, {Frenk}, {Hawkins},
  {Jackson}, {Lahav}, {Lewis}, {Lumsden}, {Maddox}, {Madgwick}, {Norberg},
  {Peacock}, {Peterson}, {Sutherland}, \& {Taylor}}]{baldry02a}
{Baldry}, I.~K., {Glazebrook}, K., {Baugh}, C.~M., {Bland-Hawthorn}, J.,
  {Bridges}, T., {Cannon}, R., {Cole}, S., {Colless}, M., {Collins}, C.,
  {Couch}, W., {Dalton}, G., {De Propris}, R., {Driver}, S.~P., {Efstathiou},
  G., {Ellis}, R.~S., {Frenk}, C.~S., {Hawkins}, E., {Jackson}, C., {Lahav},
  O., {Lewis}, I., {Lumsden}, S., {Maddox}, S., {Madgwick}, D.~S., {Norberg},
  P., {Peacock}, J.~A., {Peterson}, B.~A., {Sutherland}, W., \& {Taylor}, K.
  2002, \apj, 569, 582

\bibitem[{Bernardi {et~al.}(2002)}]{bernardi02q}
Bernardi, M. {et~al.} 2002, \aj, submitted (astro-ph/0110344)

\bibitem[{{Bessell}(1990)}]{bessell90a}
{Bessell}, M.~S. 1990, \pasp, 102, 1181

\bibitem[{Blair \& Gilmore(1982)}]{blair82a}
Blair, M. \& Gilmore, G. 1982, \pasp, 94, 742

\bibitem[{Blanton {et~al.}(2002{\natexlab{a}})Blanton, Brinkmann, Csabai, Doi,
  Eisenstein, Fukugita, Gunn, Hogg, \& Schlegel}]{blanton02b}
Blanton, M.~R., Brinkmann, J., Csabai, I., Doi, M., Eisenstein, D.~J.,
  Fukugita, M., Gunn, J.~E., Hogg, D.~W., \& Schlegel, D.~J.
  2002{\natexlab{a}}, \aj, in press (astro-ph/0205243)

\bibitem[{Blanton {et~al.}(2002{\natexlab{b}})Blanton, Lin, Lupton, Maley,
  Young, Zehavi, \& J.}]{blanton02a}
Blanton, M.~R., Lin, H., Lupton, R.~H., Maley, F.~M., Young, N., Zehavi, I., \&
  J., L. 2002{\natexlab{b}}, \aj, in press (astro-ph/0105535)

\bibitem[{Blanton {et~al.}(2001)}]{blanton01a}
Blanton, M.~R. {et~al.} 2001, \aj, 121, 2358

\bibitem[{Blanton {et~al.}(2002{\natexlab{c}})}]{blanton02e}
Blanton, M.~R. {et~al.} 2002{\natexlab{c}}, in preparation

\bibitem[{{Bruzual A.} \& {Charlot}(1993)}]{bruzual93a}
{Bruzual A.}, G. \& {Charlot}, S. 1993, \apj, 405, 538

\bibitem[{{Cole} {et~al.}(2000){Cole}, {Lacey}, {Baugh}, \& {Frenk}}]{cole00a}
{Cole}, S., {Lacey}, C.~G., {Baugh}, C.~M., \& {Frenk}, C.~S. 2000, \mnras,
  319, 168

\bibitem[{{Cole} {et~al.}(2001){Cole}, {Norberg}, {Baugh}, {Frenk},
  {Bland-Hawthorn}, {Bridges}, {Cannon}, {Colless}, {Collins}, {Couch},
  {Cross}, {Dalton}, {De Propris}, {Driver}, {Efstathiou}, {Ellis},
  {Glazebrook}, {Jackson}, {Lahav}, {Lewis}, {Lumsden}, {Maddox}, {Madgwick},
  {Peacock}, {Peterson}, {Sutherland}, \& {Taylor}}]{cole01a}
{Cole}, S., {Norberg}, P., {Baugh}, C.~M., {Frenk}, C.~S., {Bland-Hawthorn},
  J., {Bridges}, T., {Cannon}, R., {Colless}, M., {Collins}, C., {Couch}, W.,
  {Cross}, N., {Dalton}, G., {De Propris}, R., {Driver}, S.~P., {Efstathiou},
  G., {Ellis}, R.~S., {Glazebrook}, K., {Jackson}, C., {Lahav}, O., {Lewis},
  I., {Lumsden}, S., {Maddox}, S., {Madgwick}, D., {Peacock}, J.~A.,
  {Peterson}, B.~A., {Sutherland}, W., \& {Taylor}, K. 2001, \mnras, 326, 255

\bibitem[{{Colless} {et~al.}(2001){Colless}, {Dalton}, {Maddox}, {Sutherland},
  {Norberg}, {Cole}, {Bland-Hawthorn}, {Bridges}, {Cannon}, {Collins}, {Couch},
  {Cross}, {Deeley}, {De Propris}, {Driver}, {Efstathiou}, {Ellis}, {Frenk},
  {Glazebrook}, {Jackson}, {Lahav}, {Lewis}, {Lumsden}, {Madgwick}, {Peacock},
  {Peterson}, {Price}, {Seaborne}, \& {Taylor}}]{colless01a}
{Colless}, M., {Dalton}, G., {Maddox}, S., {Sutherland}, W., {Norberg}, P.,
  {Cole}, S., {Bland-Hawthorn}, J., {Bridges}, T., {Cannon}, R., {Collins}, C.,
  {Couch}, W., {Cross}, N., {Deeley}, K., {De Propris}, R., {Driver}, S.~P.,
  {Efstathiou}, G., {Ellis}, R.~S., {Frenk}, C.~S., {Glazebrook}, K.,
  {Jackson}, C., {Lahav}, O., {Lewis}, I., {Lumsden}, S., {Madgwick}, D.,
  {Peacock}, J.~A., {Peterson}, B.~A., {Price}, I., {Seaborne}, M., \&
  {Taylor}, K. 2001, \mnras, 328, 1039

\bibitem[{Cross {et~al.}(2001)}]{cross01a}
Cross, N. {et~al.} 2001, \mnras, 324, 825

\bibitem[{Davis \& Huchra(1982)}]{davis82a}
Davis, M. \& Huchra, J. 1982, \apj, 254, 437

\bibitem[{Efstathiou {et~al.}(1988)Efstathiou, Ellis, \&
  Peterson}]{efstathiou88a}
Efstathiou, G., Ellis, R.~S., \& Peterson, B.~S. 1988, \mnras, 232, 431

\bibitem[{Eisenstein {et~al.}(2001)}]{eisenstein01a}
Eisenstein, D.~J. {et~al.} 2001, accepted by \aj (astro-ph/0108153)

\bibitem[{Fan(1999)}]{fan99a}
Fan, X. 1999, \aj, 117, 2528

\bibitem[{Folkes {et~al.}(1999)Folkes, Ronen, Price, Lahav, Colless, Maddox,
  Deeley, Glazebrook, Bland-Hawthorn, Cannon, Cole, Collins, Couch, Driver,
  Dalton, Efstathiou, Ellis, Frenk, Kaiser, Lewis, Lumsden, Peacock, Peterson,
  Sutherland, \& Taylor}]{folkes99a}
Folkes, S., Ronen, S., Price, I., Lahav, O., Colless, M., Maddox, S., Deeley,
  K., Glazebrook, K., Bland-Hawthorn, J., Cannon, R., Cole, S., Collins, C.,
  Couch, W., Driver, S., Dalton, G., Efstathiou, G., Ellis, R., Frenk, C.,
  Kaiser, N., Lewis, I., Lumsden, S., Peacock, J., Peterson, B., Sutherland,
  W., \& Taylor, K. 1999, \mnras, 308, 459

\bibitem[{Fukugita {et~al.}(1996)Fukugita, Ichikawa, Gunn, Doi, Shimasaku, \&
  Schneider}]{fukugita96a}
Fukugita, M., Ichikawa, T., Gunn, J.~E., Doi, M., Shimasaku, K., \& Schneider,
  D.~P. 1996, \aj, 111, 1748

\bibitem[{Gunn {et~al.}(1998)Gunn, Carr, Rockosi, Sekiguchi,
  {et~al.}}]{gunn98a}
Gunn, J.~E., Carr, M.~A., Rockosi, C.~M., Sekiguchi, M., {et~al.} 1998, \aj,
  116, 3040

\bibitem[{{Hayes}(1985)}]{hayes85a}
{Hayes}, D.~S. 1985, in IAU Symp. 111: Calibration of Fundamental Stellar
  Quantities, Vol. 111, 225--249

\bibitem[{Hogg(1999)}]{hogg99cosm}
Hogg, D.~W. 1999, astro-ph/9905116

\bibitem[{{Huchra} {et~al.}(1983){Huchra}, {Davis}, {Latham}, \&
  {Tonry}}]{huchra83a}
{Huchra}, J., {Davis}, M., {Latham}, D., \& {Tonry}, J. 1983, \apjs, 52, 89

\bibitem[{Kurucz(1991)}]{kurucz91a}
Kurucz, R.~L. 1991, in Precision Photometry:\ Astrophysics of the Galaxy, L.
  Davis Press, Inc., 27--44

\bibitem[{Lin {et~al.}(1996)Lin, Kirshner, Shectman, Landy, Oemler, Tucker, \&
  Schechter}]{lin96a}
Lin, H., Kirshner, R.~P., Shectman, S.~A., Landy, S.~D., Oemler, A., Tucker,
  D.~L., \& Schechter, P.~L. 1996, \apj, 464, 60

\bibitem[{{Lin} {et~al.}(1999){Lin}, {Yee}, {Carlberg}, {Morris}, {Sawicki},
  {Patton}, {Wirth}, \& {Shepherd}}]{lin99a}
{Lin}, H., {Yee}, H. K.~C., {Carlberg}, R.~G., {Morris}, S.~L., {Sawicki}, M.,
  {Patton}, D.~R., {Wirth}, G., \& {Shepherd}, C.~W. 1999, \apj, 518, 533

\bibitem[{Liske {et~al.}(2002)Liske, Lemon, Driver, Cross, \& Couch}]{liske02a}
Liske, J., Lemon, D.~J., Driver, S.~P., Cross, N. J.~G., \& Couch, W.~J. 2002,
  submitted to \mnras\ (astro-ph/0207555)

\bibitem[{Maddox {et~al.}(1990)Maddox, Efstathiou, \& Sutherland}]{maddox90a}
Maddox, S.~J., Efstathiou, G., \& Sutherland, W.~J. 1990, \mnras, 246, 433

\bibitem[{Metcalfe {et~al.}(1995)Metcalfe, Fong, \& Shanks}]{metcalfe95a}
Metcalfe, N., Fong, R., \& Shanks, T. 1995, \mnras, 274, 769

\bibitem[{Norberg {et~al.}(2002)}]{norberg02a}
Norberg, P. {et~al.} 2002, \mnras, 332, 827

\bibitem[{{O'Neil} \& {Bothun}(2000)}]{oneil00a}
{O'Neil}, K. \& {Bothun}, G. 2000, \apj, 529, 811

\bibitem[{Petrosian(1976)}]{petrosian76a}
Petrosian, V. 1976, \apj, 209, L1

\bibitem[{Pier {et~al.}(2002)Pier, A., Hindsley, Hennessy, Kent, Lupton, \&
  {Ivezi{\' c}}}]{pier02a}
Pier, J.~R., A., M.~J., Hindsley, R.~B., Hennessy, G.~S., Kent, S.~M., Lupton,
  R.~H., \& {Ivezi{\' c}}, Z. 2002, \aj, submitted

\bibitem[{Richards {et~al.}(2002)}]{richards02a}
Richards, G. {et~al.} 2002, \aj, 123, 2945

\bibitem[{Schlegel {et~al.}(1998)Schlegel, Finkbeiner, \& Davis}]{schlegel98a}
Schlegel, D.~J., Finkbeiner, D.~P., \& Davis, M. 1998, \apj, 500, 525

\bibitem[{{Schneider} {et~al.}(1983){Schneider}, {Gunn}, \&
  {Hoessel}}]{schneider83a}
{Schneider}, D.~P., {Gunn}, J.~E., \& {Hoessel}, J.~G. 1983, \apj, 264, 337

\bibitem[{S\'ersic(1968)}]{sersic68a}
S\'ersic, J.~L. 1968, Atlas de Galaxias Australes (Cordoba: Obs.\ Astronomico)

\bibitem[{Shectman {et~al.}(1996)Shectman, Landy, Oemler, Tucker, Lin,
  Kirshner, \& Schechter}]{shectman96a}
Shectman, S.~A., Landy, S.~D., Oemler, A., Tucker, D.~L., Lin, H., Kirshner,
  R.~P., \& Schechter, P.~L. 1996, \apj, 470, 172

\bibitem[{Smith {et~al.}(2002)Smith, Tucker, Kent, Richmond, Fukugita,
  Ichikawa, Ichikawa, Jorgensen, Uomoto, Gunn, Hamabe, Watanabe, Tolea, Henden,
  Annis, Pier, McKay, Brinkmann, Chen, Holtzman, Shimasaku, \& York}]{smith02a}
Smith, J.~A., Tucker, D.~L., Kent, S., Richmond, M.~W., Fukugita, M., Ichikawa,
  T., Ichikawa, S.-I., Jorgensen, A.~M., Uomoto, A., Gunn, J.~E., Hamabe, M.,
  Watanabe, M., Tolea, A., Henden, A., Annis, J., Pier, J.~R., McKay, T.~A.,
  Brinkmann, J., Chen, B., Holtzman, J., Shimasaku, K., \& York, D.~G. 2002,
  \aj, 123, 2121

\bibitem[{Stoughton {et~al.}(2002)}]{stoughton02a}
Stoughton, C. {et~al.} 2002, \aj, 123, 485

\bibitem[{Strauss {et~al.}(2002)}]{strauss02a}
Strauss, M.~A. {et~al.} 2002, \aj, accepted

\bibitem[{{Wright}(2001)}]{wright01a}
{Wright}, E.~L. 2001, \apjl, 556, L17

\bibitem[{York {et~al.}(2000)}]{york00a}
York, D. {et~al.} 2000, \aj, 120, 1579

\end{thebibliography}
\end{document}